\documentclass[prd,twocolumn,showpacs,reprint,preprintnumbers,nofootinbib,amsmath,amssymb]{revtex4-2}

\RequirePackage[colorlinks=true
,urlcolor=blue
,anchorcolor=blue
,citecolor=blue
,filecolor=blue
,linkcolor=blue
,menucolor=blue
,linktocpage=true
,pdfproducer=medialab
,pdfa=true
]{hyperref}

\usepackage{amsmath,amssymb,amsthm,amsfonts}
\usepackage{graphicx,tabularx}
\usepackage{color}
\usepackage{multirow}  
\usepackage{comment}
\usepackage{slashed}
\usepackage{enumitem}
\usepackage{cleveref}
\usepackage{rotating}
\usepackage[english]{babel}
\usepackage[font=small,labelfont=bf,justification=centerlast]{caption}
\usepackage{subcaption}
\usepackage{float}

\usepackage{breqn} 
\makeatletter 
\let\cref@old@eq@setnumber\eq@setnumber 
\def\eq@setnumber{%
\cref@old@eq@setnumber%
\cref@constructprefix{equation}{\cref@result}%
\protected@xdef\cref@currentlabel{%
[equation][\arabic{equation}][\cref@result]\p@equation\theequation}} 
\makeatother

\usepackage{cases}

\crefname{section}{Sec.}{Secs.}
\crefname{figure}{Fig.}{Figs.}
\crefname{equation}{Eq.}{Eqs.}
\crefname{appendix}{Appendix}{Appendices}
\setlist[description]{leftmargin=0.4cm}
\setlist[itemize]{leftmargin=0.4cm}

\newcommand{\be}{\begin{equation}\begin{aligned}}
\newcommand{\ee}{\end{aligned}\end{equation}}

\newcommand{\beq}{\begin{equation}}
\newcommand{\eeq}{\end{equation}}
\newcommand{\beqa}{\begin{eqnarray}}
\newcommand{\eeqa}{\end{eqnarray}}

\newcommand{\kev}{\text{keV}}
\newcommand{\mev}{\text{MeV}}
\newcommand{\gev}{\text{GeV}}

\newcommand{\m}{\text{m}}

\renewcommand{\eqref}[1]{Eq.~(\ref{#1})}

\newcommand{\eg}{{\em e.g.}}

\newcommand{\TODO}[1]{\textcolor{green}{TODO}}
\RequirePackage[normalem]{ulem}

\DeclareUnicodeCharacter{2212}{\textendash}

\makeatletter
\def\l@subsubsection#1#2{}
\makeatother

\usepackage{multirow}
\usepackage{makecell}

\usepackage{fontawesome}

\begin{document}
\title{Looking forward to inelastic DM with electromagnetic form factors at FASER and beam dumps}

\author{Krzysztof Jod\l{}owski}
\email{k.jodlowski@ibs.re.kr}
\affiliation{Particle Theory and Cosmology Group\char`,{} Center for Theoretical Physics of the Universe\char`,{} Institute for Basic Science (IBS)\char`,{} Daejeon\char`,{} 34126\char`,{} Korea}

\begin{abstract}
Inelastic Dark Matter (iDM) is an interesting thermal DM scenario that can pose challenges for conventional detection methods.
However, recent studies demonstrated that iDM coupled to a photon by electric or magnetic dipole moments can be effectively constrained by intensity frontier experiments using the displaced single-photon decay signature.
In this work, we show that by utilizing additional signatures for such models, the sensitivity reach can be increased towards the short-lived regime, $\gamma c\tau \sim O(1)\,\m$, which can occur in the region of the parameter space relevant to successful thermal freeze-out. 
These processes are secondary iDM production taking place by upscattering in front of the decay vessel and electron scattering.
Additionally, we consider dimension-6 scenarios of photon-coupled iDM - the anapole moment and the charge radius operator - where the leading decay of the heavier iDM state is $\chi_1 \to \chi_0 e^+ e^-$, resulting in a naturally long-lived $\chi_1$.
We find that the decays of $\chi_1$ at FASER2, MATHUSLA, and SHiP will constrain these models more effectively than the scattering signature considered for the elastic coupling case, while secondary production yields similar constraints as the scattering. \href{https://github.com/krzysztofjjodlowski/Looking_forward_to_photon_coupled_LLPs}{\faGithub}
\end{abstract}

\maketitle

\section{\label{sec:intro}Introduction}

Dark Matter (DM) is an electrically neutral form of matter which, as a diverse set of observations indicates, makes up a significant portion of the energy content of the Universe \cite{Einasto:2009zd,vandenBergh:1999sa,Bergstrom:2000pn,Bertone:2010zza,Bertone:2016nfn}.
Arguably, one of the most motivated mechanism for DM production is its thermal freeze-out from the thermal plasma taking place in the early Universe \cite{Vysotsky:1977pe,Lee:1977ua}. 
Null searches for Weakly Interacting Massive Particles (WIMPs) \cite{Roszkowski:2017nbc,Arcadi:2017kky}, which would acquire their relic density in this way, motivates the exploration of more elaborate scenarios.

An interesting proposal that shares many desirable features with WIMPs is the inelastic DM (iDM) \cite{Tucker-Smith:2001myb}. In its original form, it assumes that the Dark Sector (DS) consists of two states that are charged under a SM or dark $U(1)$ gauge symmetry, and the coupling between the DS species and the gauge boson is non-diagonal.
Moreover, although DM needs to be electrically neutral (it can be at most millicharged) \cite{Davidson:2000hf,McDermott:2010pa}, interactions with the photon can occur through electromagnetic (EM) form factors \cite{Pospelov:2000bq,Sigurdson:2004zp,Gardner:2008yn,Masso:2009mu,Barger:2010gv}.

Combining these two ideas, the iDM scenario can be realized by introducing such a higher dimensional EM operator, \eg, an electric or magnetic dipole moment (EDM/MDM) \cite{Masso:2009mu,Chang:2010en,Baumgart:2009tn,Weiner:2012gm,Weiner:2012cb,Cline:2012bz,Izaguirre:2015zva,Barducci:2022gdv}, while preserving the distinct inelastic character of the scattering signatures at collider or direct detection searches, as in the original scenario \cite{Tucker-Smith:2001myb}.
One consequence of such coupling is the decay of the heavier iDM state into the lighter DS state and SM particles. 
If the masses of the iDM states are within $\sim\,$sub-GeV range, and the mass splitting between them is small - incidentally, this regime is usually needed to obtain the correct relic density - the heavier state is a long-lived particle (LLP). 
In fact, the iDM coupled to a dark photon is one of benchmarks for intensity frontier experiments looking for highly displaced LLP decays \cite{Battaglieri:2017aum,Beacham:2019nyx}.

In particular, recent work has shown \cite{Dienes:2023uve} that intensity frontier experiments such as the current FASER \cite{Feng:2017uoz,Feng:2017vli} and the upcoming FASER2 \cite{FASER:2018ceo,FASER:2018bac,FASER:2021ljd} and SHiP \cite{SHiP:2015vad,Alekhin:2015byh} detectors will cover much of the allowed parameter space for iDM connected to the SM via EDM or MDM.
We note that the elastic coupling scenario was studied first, see, \eg, \cite{Chu:2018qrm,Chu:2020ysb,Kling:2022ykt}, but in this case the DM species is stable and therefore only its scattering signatures can be used.
Moreover, since the LLP decays taking place inside a detector can be more efficient than scattering with electrons, it was found that the searches for decays of the heavier DS species, $\chi_1 \to \chi_0 \gamma$, will cover substantially larger part of the parameter space than in the case of elastic coupling \cite{Dienes:2023uve}.

The presence of extended DS, that is two species with small mass splitting between them, allows one to search not only for LLP decays, but it also enables secondary LLP production \cite{Jodlowski:2019ycu,Jodlowski:2020vhr,Jodlowski:2023sbi}. 
It takes place just in front of the decay vessel by upscattering of the stable DS species into the heavier one.

In this work, we extend \cite{Dienes:2023uve} by considering secondary LLP production followed by its rapid decay and the electron scattering signature for EDM/MDM iDM. 
In addition, we consider additional scenarios for the iDM - when it is connected to the SM via anapole moment (AM) or charge radius (CR) operator. In these cases, the leading decay is phase-space suppressed, $\chi_1 \to \chi_0 e^+ e^-$, which could naturally explain the long-lived nature of $\chi_1$. We also study the impact of secondary LLP production and scattering with electrons.

The paper is organized as follows.
In \cref{sec:models}, we introduce the scenarios for iDM connection to a photon via mass-dimension five and six operators.
We review their properties, and in particular discuss the parametric dependence of LLP decay length, which can be probed at the intensity frontier.
In \cref{sec:relic_density}, we discuss the relic density of $\chi_0$ obtained by thermal freeze-out. We identify the key annihilation/co-annihilation processes entering the Boltzmann equations, and we discuss their solutions.
In \cref{sec:LLP_sign}, we discuss experimental signatures of long-lived iDM: displaced decays, secondary LLP production, and electron scattering in various experiments. We also provide details of our simulation.
In \cref{sec:results}, we discuss our results, in particular the sensitivity plots of past and future experiments for the four iDM scenarios considered.
We also compare our results with previous works.
We conclude in \cref{sec:conclusions}, while expressions for the decay widths and cross sections used in our analysis were relegated to \cref{app:prod,app:decays,app:e_scat,app:relic}.

\section{Models\label{sec:models}}

We consider iDM composed of two dark fermions - $\chi_0$, $\chi_1$ - coupled to a photon by two dimension-5 operators, EDM and MDM, and by two dimension-6 operators, AM and CR operator.
The key property of this scenario is that the DS states have split masses, $\Delta \equiv (m_{\chi_1}-m_{\chi_0})/m_{\chi_0}>0$.

The interactions are described by the following effective Lagrangian \cite{Dienes:2023uve,Chu:2018qrm,Chu:2020ysb,Kling:2022ykt}:\footnote{For MDM (EDM) elastic case, a different parametrization using $\mu_{\mathrm{M}}$ ($d_{\mathrm{E}}$) is commonly used, where $\Lambda_{\mathrm{M}} = 2/\mu_{\mathrm{M}}$ ($\Lambda_{\mathrm{E}} = 2/d_{\mathrm{E}}$).}
\begin{dmath}[labelprefix={eq:}]
  \!\!\mathcal{L} \supset \frac{1}{\Lambda_{\mathrm{E}}} \bar{\chi}_1 \sigma^{\mu \nu} \gamma^5 \chi_0 F_{\mu \nu} + \frac{1}{\Lambda_{\mathrm{M}}} \bar{\chi}_1 \sigma^{\mu \nu} \chi_0 F_{\mu \nu} - a_\chi \bar{\chi}_1 \gamma^{\mu} \gamma^5 \chi_0\, \partial^\nu F_{\mu \nu} + b_\chi \bar{\chi}_1 \gamma^{\mu} \chi_0\, \partial^\nu F_{\mu \nu},
  \label{eq:lagr}
\end{dmath}
where $1/\Lambda_{\mathrm{E}}$ and $1/\Lambda_{\mathrm{M}}$ are couplings of mass-dimension -1, while $a_\chi$ and $b_\chi$ are couplings of mass-dimension -2; $\sigma^{\mu\nu}=\frac i2[\gamma^\mu,\gamma^\nu]$, and $F_{\mu\nu}$ is the EM field strength tensor.
The first line describes EDM and MDM \cite{Masso:2009mu,Chang:2010en,Baumgart:2009tn,Weiner:2012gm,Weiner:2012cb,Cline:2012bz,Izaguirre:2015zva}, while the second line describes AM and CR operator \cite{Ho:2012bg,Gao:2013vfa,DelNobile:2014eta,Alves:2017uls}.

This Lagrangian is an adaptation of the EFT framework used to describe fermionic DM with an elastic coupling to a photon \cite{DelNobile:2011uf,Kavanagh:2018xeh} to iDM, keeping operators up to mass-dimension 6.
We note that considering, \eg, scalar iDM is also possible, which we leave for further work.

Moreover, \cref{eq:lagr} is only an effective description valid for energies below the electroweak breaking scale, and in general requires UV completion, see, \eg, \cite{Weiner:2012gm,Weiner:2012cb} for a discussion of such a construction for magnetic iDM and a dimension-7 Rayleigh operator.
In fact, as discussed in \cite{Masso:2009mu,Chang:2010en,Izaguirre:2015zva}, \cref{eq:lagr} may arise from the elastic couplings case, where a Dirac fermion $\chi$ also obtains a Majorana mass, leading to a small mass splitting $\Delta$. 
In such framework, another UV completion mechanisms have been discussed, \eg, \cite{Foadi:2008qv,Bagnasco:1993st,Antipin:2015xia,Raby:1987ga,Pospelov:2008qx}.
Finally, replacing $F_{\mu \nu}$ with $B_{\mu \nu}$, the hypercharge field strength tensor, substantially modifies phenomenology at, \eg, the LHC for $m_\chi \gtrsim 100\,\gev$ \cite{Arina:2020mxo}.
Since we will be discussing a mass range of $1\,\mev \lesssim \m_{\chi_0} \lesssim 10\,\gev$, as well as the small momentum exchange regime in scattering processes involving the iDM species, \cref{eq:lagr} is an adequate description, except for interpreting LEP and LHC limits, which can depend on specifics of the EW completion and EFT validity at the hard scales of those searches.

In this mass range, and for a small mass splitting, $\Delta \ll 1$, the decay width of $\chi_1$ is suppressed, and $\chi_1$ acts as a LLP.
Interestingly, the same regime is needed for successful thermal freeze-out of $\chi_0$ from the thermal plasma in the early Universe thanks to the co-annihilation processes.
This behavior is typical in iDM scenarios \cite{Tucker-Smith:2001myb}, and is discussed at length within our setup in \cref{sec:relic_density}.

The lifetime of $\chi_1$ strongly depends on the nature and dimensionality of the interactions.
In particular, for EDM/MDM it is determined by two-body decays, $\chi_1 \to \chi_0 \gamma$, with widths given by \cref{eq:Gamma_EDM_MDM}.
For dimension-6 operators, such two-body decays are not possible because the $\chi_0{\text -}\chi_1{\text -}\gamma$ coupling depends on the four-momentum of the photon, which vanishes on-shell. 
On the other hand, three-body decays of $\chi_1$ into $\chi_0$ and a pair of charged leptons are allowed because the photon four-momentum from the aforementioned coupling is canceled by the propagator of the off-shell photon.

Therefore, $\chi_1 \to \chi_0 e^+ e^-$, described by \cref{eq:Gamma_AM_CR}, is the leading decay channel.
Due to phase-space suppression, the $\chi_1$ lifetime that is relevant for the intensity frontier searches in the AM/CR scenario is shifted towards larger values of $m_{\chi_0}$ and $\Delta$ compared to the EDM/MDM scenario.
Moreover, we checked that this process does not lead to a noticeable injection of positrons relevant to the $511 \,\kev$ excess in either decaying or excited DM scenarios (see discussion in the recent work \cite{Cappiello:2023qwl} and references therein), since the $\chi_1$ abundance (from freeze-out or upscattering $\chi_0\chi_0 \to \chi_1\chi_1$) is negligible in the MeV-GeV mass regime we investigate.

Given the intensity frontier detectors are typically situated approximately $100\,\m$ away from the LLP production point, they can effectively probe typical LLP decay lengths of
\begin{widetext}
  \be
    \text{EDM/MDM:\quad}  \
    &   d_{\chi_1} \simeq 100 \,\m \times \left(\frac{E}{1000\,\gev}\right) \left(\frac{0.1\,\gev}{m_{\chi_0}}\right)^4 \left(\frac{0.05}{\Delta}\right)^3 \left(\frac{\Lambda}{2.84 \times 10^{3}\,\gev}\right)^2, \\
    \text{AM:\quad} \    
    &   d_{\chi_1} \simeq 100 \,\m \times \left(\frac{E}{1000\,\gev}\right) \left(\frac{1\,\gev}{m_{\chi_0}}\right)^4 \left(\frac{0.05}{\Delta}\right)^5 \left(\frac{6.53 \times 10^{-3}\,\gev^{-2}}{a_\chi}\right)^2, \\
    \text{CR:\quad}  \   
    &   d_{\chi_1} \simeq 100 \,\m \times \left(\frac{E}{1000\,\gev}\right) \left(\frac{1\,\gev}{m_{\chi_0}}\right)^4 \left(\frac{0.05}{\Delta}\right)^5 \left(\frac{1.13\times 10^{-2}\,\gev^{-2}}{b_\chi}\right)^2,
    \label{eq:dbar}
  \ee
\end{widetext}
where $E$ is the energy of $\chi_1$ the in the LAB frame, and we fixed all the parameters except the coupling constants to the typical values that can be covered by FASER2.

We note that by using the above $d_{\chi_1}$ formula for EDM/MDM case, we find disagreement with Eq. 2.3 of \cite{Dienes:2023uve}, which underestimates the lifetime by a factor of 260. 
Looking at the results shown in the bottom panels of Fig. 3 therein, the benchmark from Eq. 2.3 lies in the lower left part of the FASER2 sensitivity curve.
However, this part of the sensitivity curve is determined by LLP species that are on the verge of being too long-lived to decay inside the detector, and therefore correspond to $d_{\chi_1} \gg 600\,\m$.
In the opposite regime, LLPs with a typical decay length $d_{\chi_1} \lesssim L$, where $L$ is the distance from their point of production to their decay ($L \sim 480\,\m$ for FASER) are in the upper right region of the sensitivity curve.
Points above this line are too short-lived to be probed, since the probability of LLP decays taking place inside a distant detector is exponentially suppressed for $d_{\chi_1} \ll L$ \cite{Bauer:2018onh,Feng:2017uoz}.
Therefore, the parameters used in Eq. 2.3 correspond to a much longer-lived LLP than indicated in \cite{Dienes:2023uve}.
On the other hand, we believe that this discrepancy is a typographical error that does not impact any of their other findings. As discussed in \cref{sec:results}, we obtained similar sensitivity reaches for the EDM/MDM cases, which is described in that section's discussion.

\section{Relic density\label{sec:relic_density}}

\begin{figure*}[tb]
  \centering
  \includegraphics[width=0.55\textwidth]{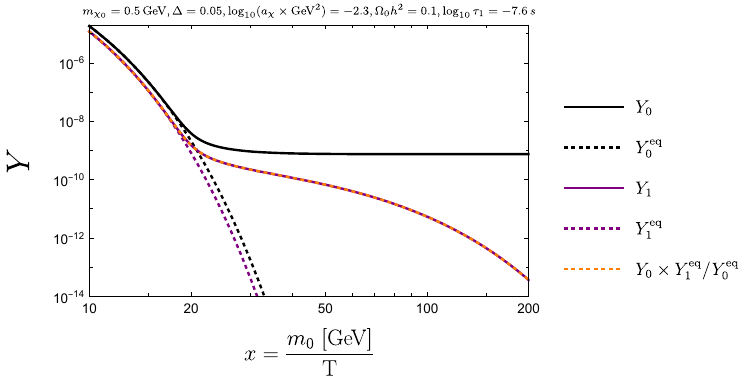}
  \includegraphics[width=0.405\textwidth]{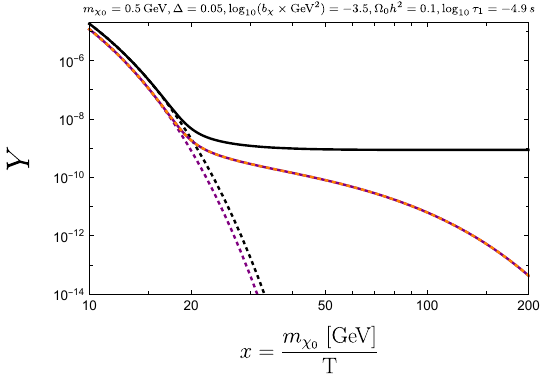}\vspace*{0.15cm}
  \includegraphics[width=0.55\textwidth]{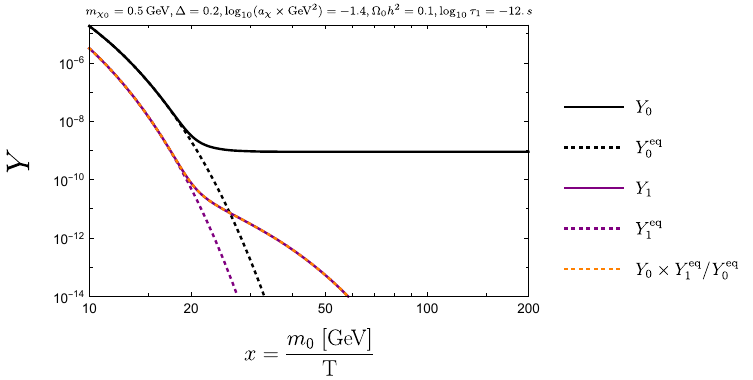}
  \includegraphics[width=0.405\textwidth]{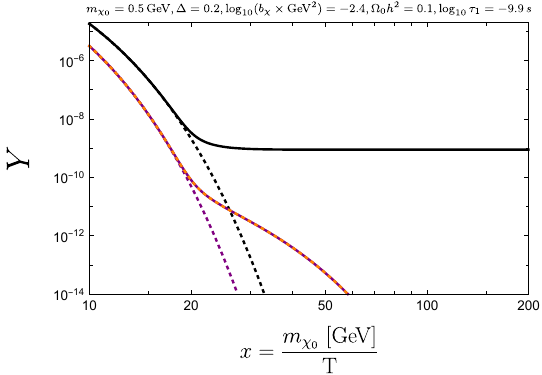}
  \caption{ 
    Solutions of the Boltzmann equations shown for $\Delta=0.05$ (top) and $\Delta=0.2$ (bottom), with a fixed $m_0 = 0.5\,\gev$, resulting in the correct relic density $\Omega_0 h^2 \simeq 0.1$. 
    Results for the AM are on the \textit{left}, while the results for the CR operator are on the \textit{right}.
    The title of each plot contains the values of the parameters used to obtain each solution and the logarithm of the lifetime of $\chi_1$. 
    The solid lines correspond to the numerical solutions of \cref{eq:boltzmann_three}, while the dashed lines denote the equilibrium comoving yields. 
    The orange dashed line indicates the relation given by \cref{eq:Y1_asympt}, which is followed by $Y_1$ over the shown range.
  }
  \label{fig:relic_density_benchmarks}
\end{figure*}

Since the photon connects the visible and dark sectors, the stable iDM species $\chi_0$ is a natural thermal DM candidate \cite{Tucker-Smith:2001myb,Masso:2009mu,Weiner:2012cb}.
The relic densities of AM/CR iDM\footnote{The relic density of EDM/MDM iDM is discussed in \cite{Dienes:2023uve}. The main distinction between these cases lies in the presence of additional annihilation channels, specifically $\chi_i \chi_i \to \gamma \gamma$, which vanish in the AM/CR scenario.} is obtained by solving the following Boltzmann equations that describe the temperature evolution of the comoving yields $Y_i \equiv n_i/s$ for $i \in \{\chi_0, \chi_1\} \equiv \{0,1\}$:
\begin{widetext}
  \begin{align}
    \label{eq:boltzmann_three}
    \frac{d Y_0}{d x} & = -\lambda \left(Y_0 Y_1 - Y_0^{\mathrm{eq}} Y_1^{\mathrm{eq}} \right) \left\langle \sigma_{0 1 \rightarrow \mathrm{SM}\mathrm{SM}} v \right\rangle -\lambda \left(Y_0 - Y_1 \frac{Y_0^{\mathrm{eq}}}{Y_1^{\mathrm{eq}}} \right) \left\langle \sigma_{0 e^- \rightarrow 1 e^-} v \right\rangle + \tilde{\lambda} \left(Y_1 - Y_0 \frac{Y_1^{\mathrm{eq}}}{Y_0^{\mathrm{eq}}} \right) \left\langle \Gamma_{1 \rightarrow 0 e^+ e^-} \right\rangle, \\
    \frac{d Y_1}{d x} & = -\lambda \left(Y_0 Y_1 - Y_0^{\mathrm{eq}} Y_1^{\mathrm{eq}} \right) \left\langle \sigma_{0 1 \rightarrow \mathrm{SM}\mathrm{SM}} v \right\rangle +\lambda \left(Y_0 - Y_1 \frac{Y_0^{\mathrm{eq}}}{Y_1^{\mathrm{eq}}} \right) \left\langle \sigma_{0 e^- \rightarrow 1 e^-} v \right\rangle - \tilde{\lambda} \left(Y_1 - Y_0 \frac{Y_1^{\mathrm{eq}}}{Y_0^{\mathrm{eq}}} \right) \left\langle \Gamma_{1 \rightarrow 0 e^+ e^-} \right\rangle, \nonumber \\
    \text{wh} &\text{ere } \lambda \equiv \sqrt{\frac{\pi}{45}} \frac{m_0 m_{\textrm{Pl.}}}{x^2} \frac{g_{*s}(m_0)}{\sqrt{g_{*}(m_0)}}\,, \hspace{0.3cm} \tilde{\lambda} \equiv \sqrt{\frac{45}{4 \pi^3}} \frac{m_{\textrm{Pl.}}}{m_0 ^2} \frac{x}{\sqrt{g_{*}(m_0)}}\,, \hspace{0.3cm} Y_{i}^{\mathrm{eq}}(x)=\frac{g_i}{g_{*s}(x)} \frac{45}{4 \pi^{4}} (r_i\,x)^{2} K_{2}[r_i\,x]\,. \nonumber
  \end{align}
\end{widetext}
We used: $x = m_{0}/T$, $r_i=m_i/m_{0}$, $m_{\textrm{Pl.}} = 1.9 \times 10^{19}\,\gev$ is the Planck mass, and $g_{*s}(T)$ and $g_{*}(T)$ are the effective number of degrees of freedom for the entropy and energy densities at temperature $T$ of the SM-DS plasma, respectively.
The brackets indicate the thermal average, \eg, $\left\langle \Gamma_{1 \rightarrow 0 e^+ e^-} \right\rangle = K_1\left(m_1 / T\right)/K_2\left(m_1 / T\right)\, \Gamma_{1 \rightarrow 0 e^+ e^-}$, where $K_n(z)$ is the modified Bessel function of the second kind \cite{NIST:DLMF}.
When relevant, we treat other decay channels analogously.

We numerically solve \cref{eq:boltzmann_three} using partial wave expansion \cite{Gondolo:1990dk} of the thermally averaged cross sections given by \cref{eq:sigmav_ll,eq:sigmav_ll_C,eq:sigmav_hadrons,eq:sigmav_gammagamma,eq:sigmav_gammagamma_C,eq:sigmav_chi1chi1_chi0chi0,eq:sigmav_chi1chi1_chi0chi0_C,eq:sigmav_chi1e_chi0e,eq:sigmav_chi1e_chi0e_C}.
We obtain the present-day yields of each species $i$, $Y_{i,0}$, which are used to compute the resulting relic density $\Omega_{i} h^2=(\rho_{i}/\rho_{\mathrm{crit}})\,h^2=(s_0\,Y_{i,0}\,m_{i}/\rho_{\mathrm{crit}})\,h^2$, where $s_0$ is the present-day entropy density.
In all four iDM scenarios, the dominant process responsible for the successful thermal freeze-out is $\chi_0$-$\chi_1$ co-annihilation into SM leptons and hadrons, $\left\langle \sigma_{0 1 \rightarrow \mathrm{SM}\,\mathrm{SM}}\, v \right\rangle$.
Other processes, in particular $\chi_1 \chi_1 \to \chi_0 \chi_0$ annihilation, are suppressed, this one by the fourth power of the coupling constant, and do not play a significant role in the thermal freeze-out.

As is typical for the co-annihilation mechanism \cite{Griest:1990kh}, the correct relic density, $\Omega_0 h^2 \simeq 0.1$, can only be obtained for sufficiently small values of mass splitting, $\Delta \lesssim O(0.5)$.
This is precisely the regime of long-lived $\chi_1$, which allows to probe such iDM scenarios at the intensity frontier \cite{Izaguirre:2015zva,Berlin:2018jbm,Dienes:2023uve}.
This bound holds because in \cref{eq:boltzmann_three} the term corresponding to $\chi_0$-$\chi_1$ co-annihilations into leptons and hadrons is suppressed by a factor of $\sim \exp(-x \cdot \Delta)$ compared to the same process with $\Delta = 0$. 
This suppression is due to the magnitude of the comoving yield $Y_1$ with respect to $Y_0$; note that the asymptotic behavior of the Bessel function at $z\to\infty$ is $K_2(z) \sim \sqrt{\pi/(2z)}\,\exp({-z})$ \cite{NIST:DLMF}.

For the EDM/MDM case, we recover the results of \cite{Dienes:2023uve}. Therefore, in this discussion, we will focus solely on the AM/CR scenarios.
In \cref{fig:relic_density_benchmarks}, we show the solutions of \cref{eq:boltzmann_three} for AM on the left and the solutions for CR on the right. 
We consider two mass splitting patterns, which will be discussed in more detail in \cref{sec:results}, corresponding to $\Delta=0.05$ (top) and $\Delta=0.2$ (bottom).

In both cases, iDM species decouple from the thermal plasma when the co-annihilation rate falls below the Hubble rate, which occurs near $x \approx 20$.
Note that the value of the coupling constant required to obtain the correct relic density is larger for the AM than for the CR operator due to the p-wave suppression of the co-annihilation cross section in the former scenario.

Moreover, similar to the case of EDM/MDM discussed in \cite{Dienes:2023uve}, once iDM departs from thermal equilibrium, the up- and down-scattering processes, as well as the $\chi_1$ decays/inverse decays, ensure the validity of the following relation: 
\be
  Y_1(x) \simeq Y_0(x)\times \frac{Y^{\mathrm{eq}}_1(x)}{Y^{\mathrm{eq}}_0(x)}.
  \label{eq:Y1_asympt}
\ee
This relation holds because in such co-annihilation freeze-out scenario, chemical and kinetic equilibrium between iDM species is preserved even after the decoupling, thus then $d(Y_0+Y_1)/dx \approx 0$, and since $Y_1 \ll Y_0$, one obtains \cref{eq:Y1_asympt}.

\begin{figure}[tb]
  \centering
  \includegraphics[width=0.425\textwidth]{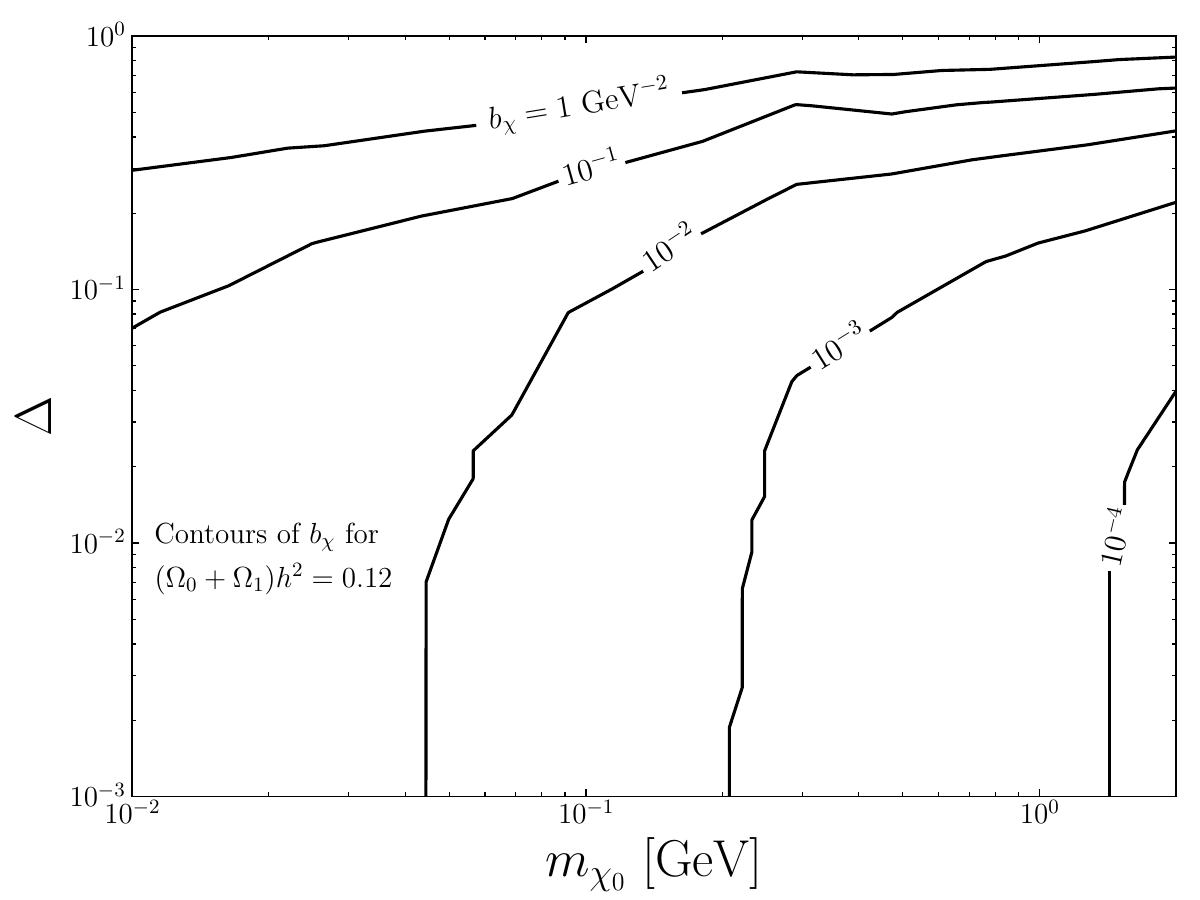}
  \caption{
  Contours of $b_\chi$ corresponding to the observed DM relic density shown in the $m_{\chi_0}$-$\Delta$ plane for the CR operator iDM. For anapole moment iDM, the resulting couplings are correspondingly larger, due to p-wave suppression.
  }
  \label{fig:CR_mchi0_Delta_scan}
\end{figure}

\section{Long-lived regime of iDM\label{sec:LLP_sign}}
In this section, we describe the iDM production mechanisms and signatures probing the long-lived nature of $\chi_1$ for the iDM scenarios introduced in \cref{sec:models}.

\subsection{Intensity frontier searches for iDM \label{sec:iDM_production}}

\paragraph{Primary LLP production}
The production rates of iDM at FASER2 have been studied and discussed in previous works.
For the elastic version of interactions given by \cref{eq:lagr}, we refer to the discussion around Fig. 1 in \cite{Kling:2022ykt}, while for EDM/MDM iDM, we refer to \cite{Dienes:2023uve} - Fig. 1 and the discussion there.

As discussed in these works (see also \cite{Chu:2020ysb}), the main channel for the production of $\overline{\chi}_0$-$\chi_1$ pairs in the mass range $m_{\chi_0} \simeq m_{\chi_1} \sim \mev$-$\gev$, which is relevant to the $\chi_1$ long-lived regime, is vector meson decays, $V \!\to\! \overline{\chi}_0 \chi_1$.
Away from kinematic thresholds, the branching ratios relative to $V\to e^+e^-$ scale as $M^2$ for the dimension-5 dipole operators~\cite{Chu:2020ysb,Dienes:2023uve} and as $M^4$ for the dimension-6 AM and CR operators, at fixed couplings. 
Heavier vector mesons can therefore provide substantial contributions if sufficiently abundant.
The formulas for the branching ratio for all iDM operators are provided in \cref{app:prod}, see \cref{eq:brV}. We also consider the next-leading contribution to the LLP production, which are three-body pseudoscalar meson decays, $P \!\to\! \gamma \overline{\chi}_0 \chi_1$.
The double differential form of the branching ratio, convenient for Monte Carlo simulation, is given by \cref{eq:br2dq2dcostheta}.

For both vector and pseudoscalar meson decays, we recovered results of \cite{Dienes:2023uve} for EDM/MDM, and in the elastic limit ($\Delta \to 0$) of all four iDM operators, we found agreement with the results from \cite{Kling:2022ykt}, except for the pseudoscalar decay for MDM.
In this case, we get the opposite sign in front of the $\sin^2\theta$ term (using their parameterization from Eq. A1 therein). 
We verified that in the $m_{\chi_1} = m_{\chi_0}\to 0$ limit, when $\frac{d{\rm BR}_{P \rightarrow \gamma \bar{\chi}_0 \chi_1}}{dq^2 d\cos\theta}$ is proportional to the $\sin^2\theta$ term, the integrated formula from \cite{Kling:2022ykt} yields a negative number.
We note that this potential discrepancy does not affect any other results from \cite{Kling:2022ykt} (or ours), since this decay is subdominant with respect to the vector meson decays.

\paragraph{Displaced LLP decays}
After the iDM species are produced, the heavier of the two, $\chi_1$, decays into $\chi_0$ and a photon (a $e^+ e^-$ pair) via the EDM/MDM (AM/CR).
The main signature considered in this work involves such decays taking place inside a dedicated detector, which is positioned at a significant distance ($\sim 100 \,\m$) from the point of production of iDM species to mitigate the SM background.
The corresponding probability of such an event is given by \cite{Bjorken:1988as}
\be
  p(E) = e^{-L/d(E)}-e^{-(L+\Delta)/d(E)},
  \label{eq:p_prim}
\ee
where $d(E)$ represents $\chi_1$ decay length in the laboratory reference frame, while $L$ indicates the distance between the start of the detector, which has length $\Delta$, and the LLP production point. 
As mentioned in the discussion following \cref{eq:dbar}, this probability is exponentially suppressed for LLPs characterized by insufficient decay length, $d \ll L$, while in the opposite regime, $d \gg L$, the suppression is only linear, $p(E) \simeq \Delta/d$ \cite{Feng:2018pew}.
Hence, $L$ determines the minimal length scale for $d$ that can be studied by such method.

Intensity frontier searches looking for $\sim\,$sub-GeV LLPs are typically only sensitive to decays to a pair of charged SM fermions or a pair of photons \cite{Battaglieri:2017aum,Beacham:2019nyx,Alimena:2019zri}.
Decays into a single high-energy photon are more challenging due to the significantly larger background compared to decays into $e^+ e^-$ or $\gamma \gamma$.
However, current LHC far-forward detector FASER \cite{Feng:2017uoz,Feng:2017vli,FASER:2022hcn} or beam dump experiments such as, \eg, the future SHiP \cite{SHiP:2015vad,Alekhin:2015byh} will have sensitivity to such signatures.
In fact, several studies, \eg, \cite{Jodlowski:2020vhr,deNiverville:2019xsx,Dienes:2023uve,Jodlowski:2023sbi}, have already investigated various BSM scenarios with LLP decaying into a single photon.

\begin{figure*}[tb]
  \centering
  \includegraphics[width=0.435\textwidth]{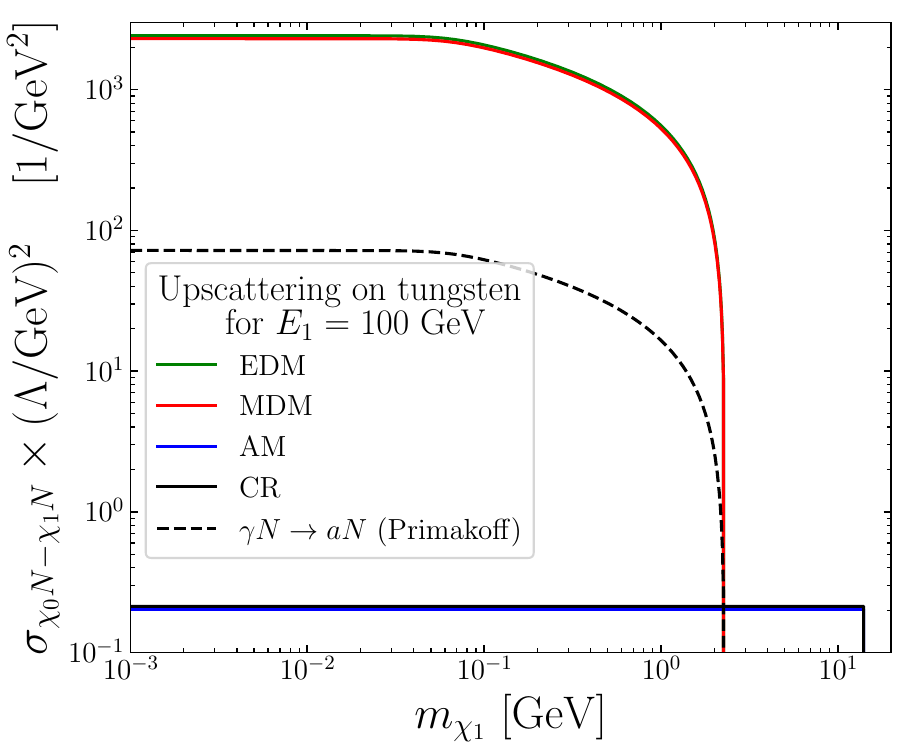}\hspace*{1.0cm}
  \includegraphics[width=0.435\textwidth]{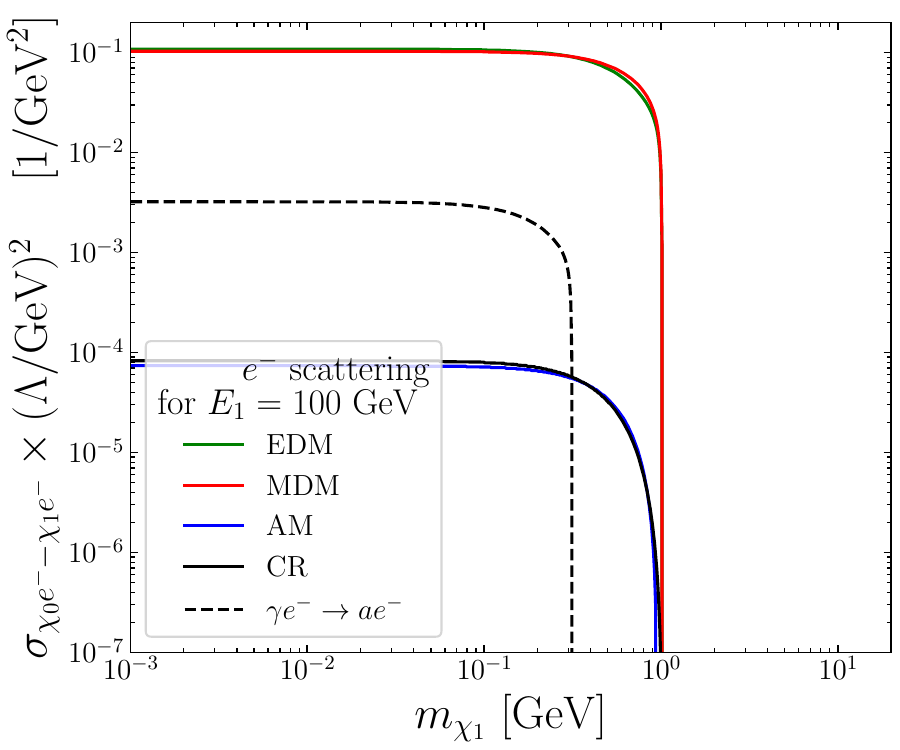}\vspace*{0.15cm}
  \includegraphics[width=0.435\textwidth]{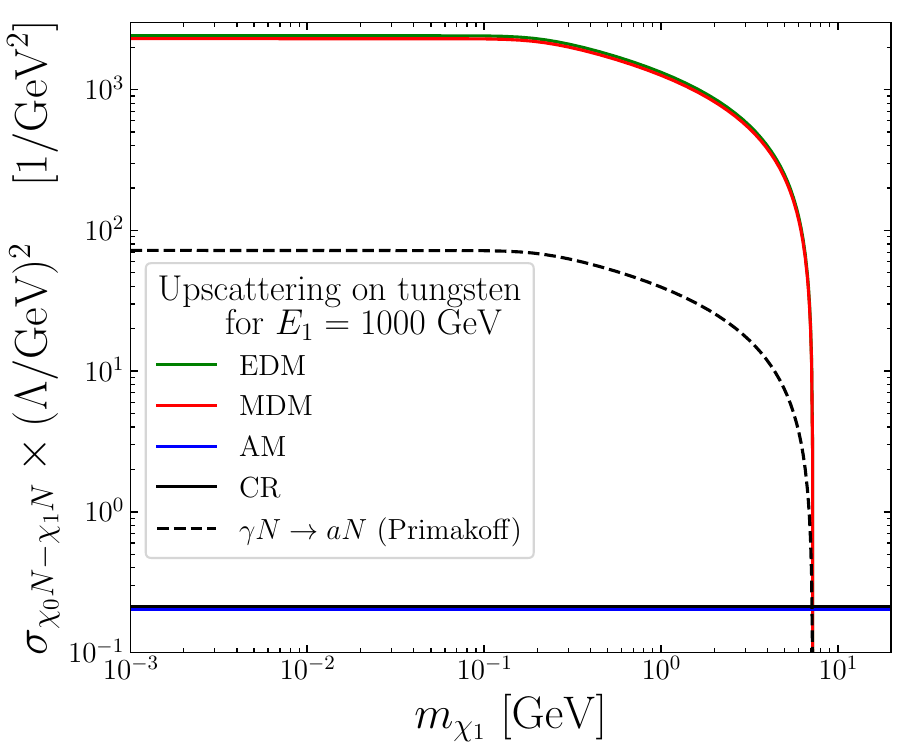}\hspace*{1.0cm}
  \includegraphics[width=0.435\textwidth]{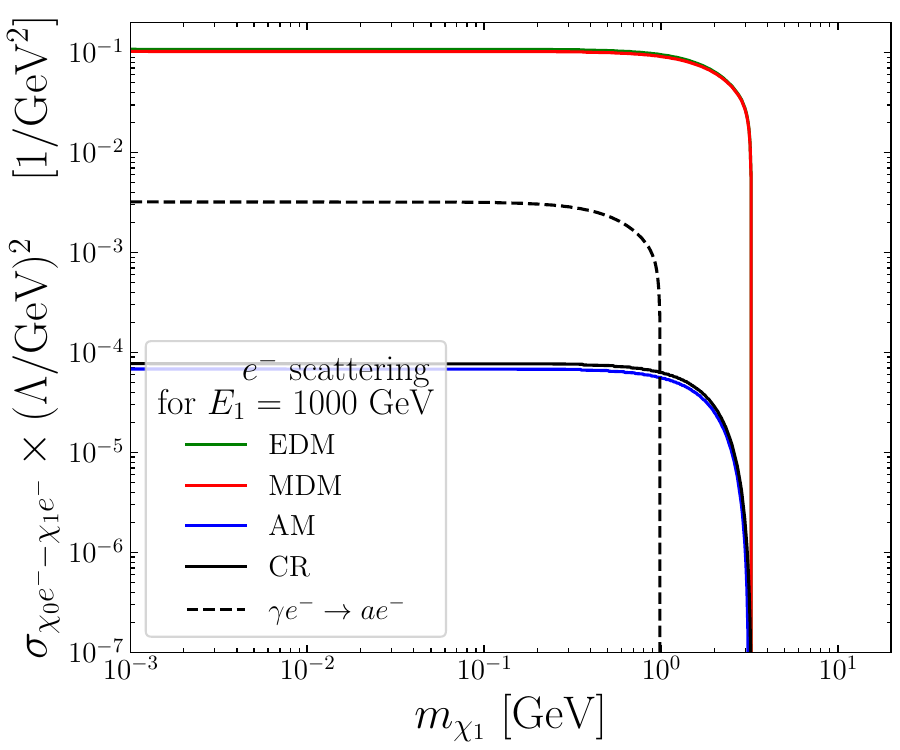}
  \caption{
    Mass dependence of the cross sections for upscattering taking place on tungsten (left) and electron scattering (right). 
    The upper panels correspond to $E_1=100\,\gev$ of the incoming particle, $\chi_0$, while the lower panels correspond to $E_1=1000\,\gev$. Note the significant suppression of the cross sections for both processes for dimension-6 operators relative to dimension-5 operators due to the lack of $1/t$ amplitude enhancement.
    Since the cross sections given in \cref{eq:Prim_iDM} overlap for both EDM/MDM and AM/CR, we shifted the overlapping lines by $5\%$. 
    For comparison, we also show the cross section for Primakoff upscattering for an axion-like particle (ALP) coupled to two photons.
    }
  \label{fig:Prim_e}
\end{figure*}

\begin{figure*}[tb]
  \centering
  \includegraphics[width=0.49\textwidth]{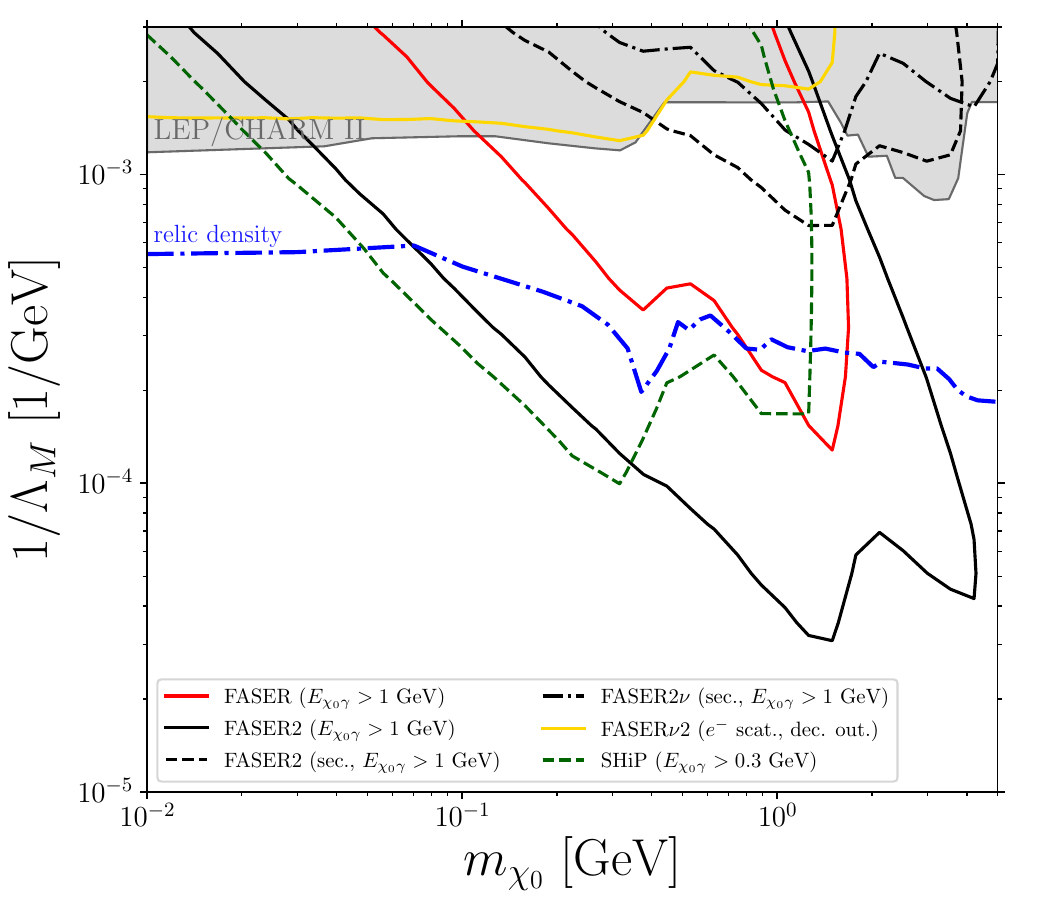}\hspace*{0.4cm}
  \includegraphics[width=0.49\textwidth]{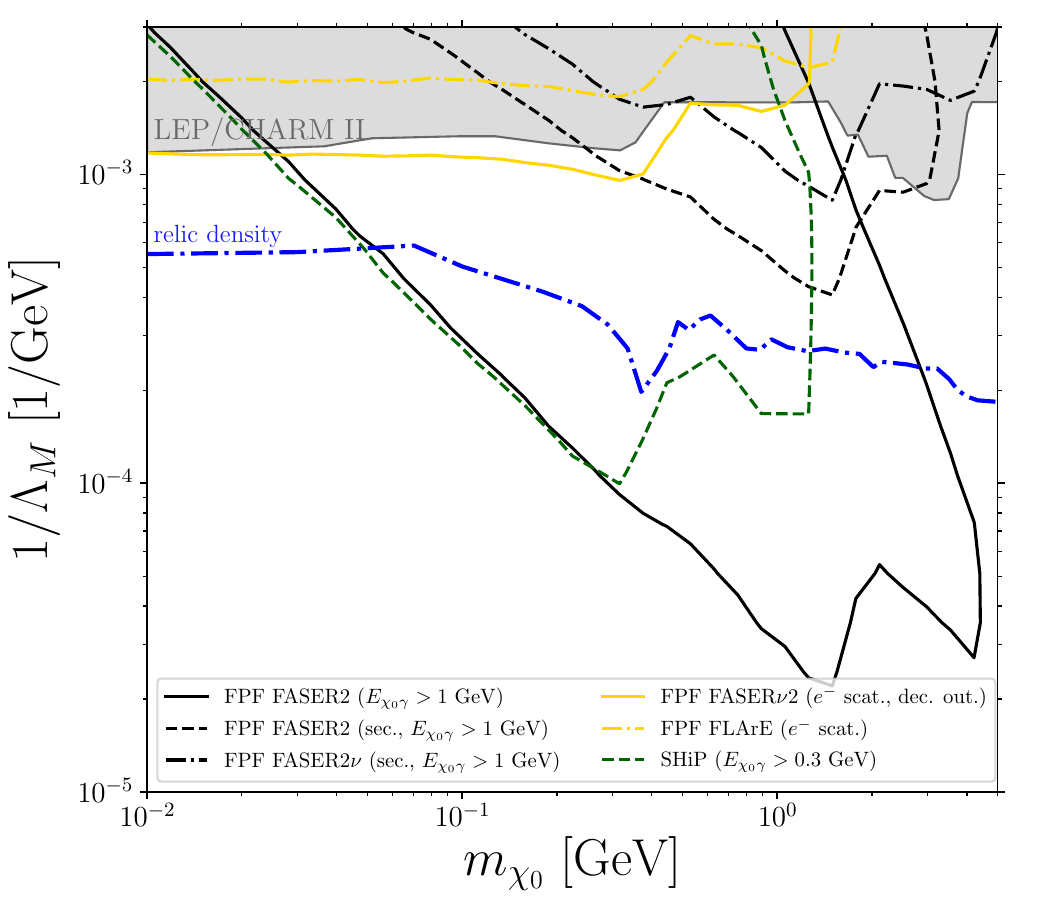}\vspace*{0.2cm}
  \includegraphics[width=0.49\textwidth]{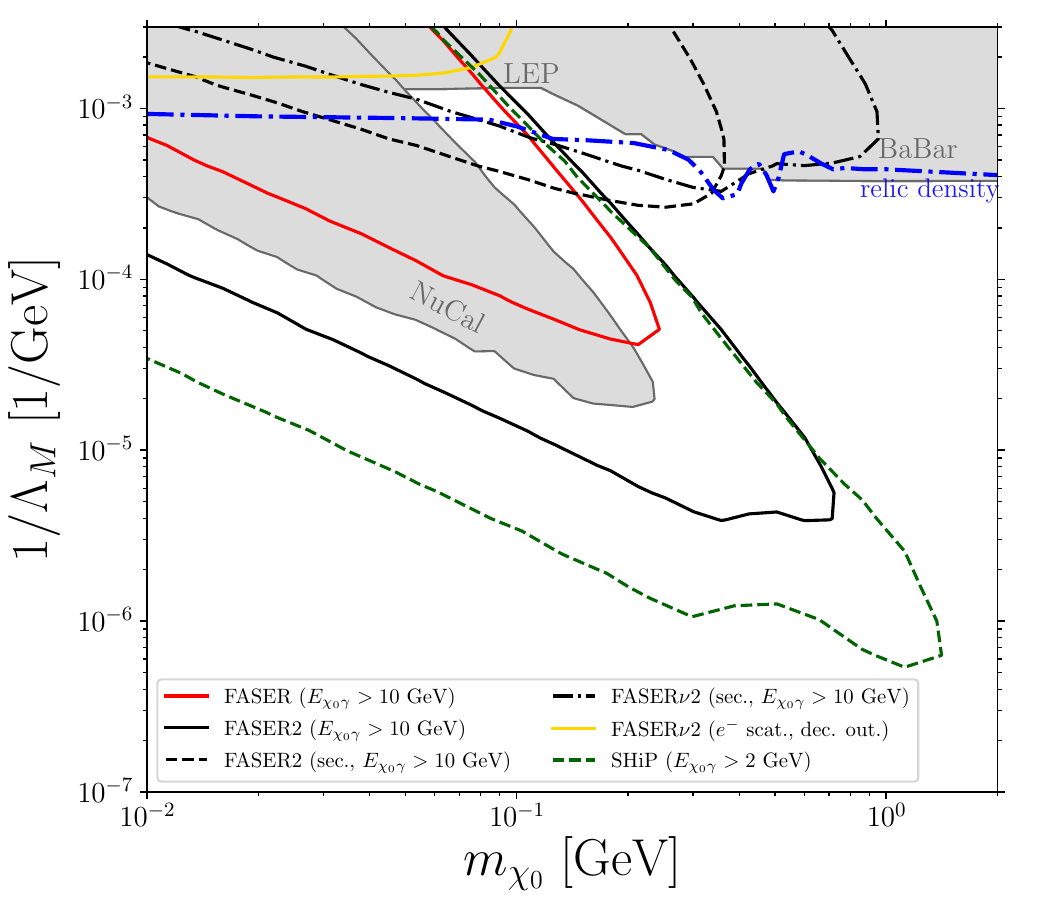}\hspace*{0.4cm}
  \includegraphics[width=0.49\textwidth]{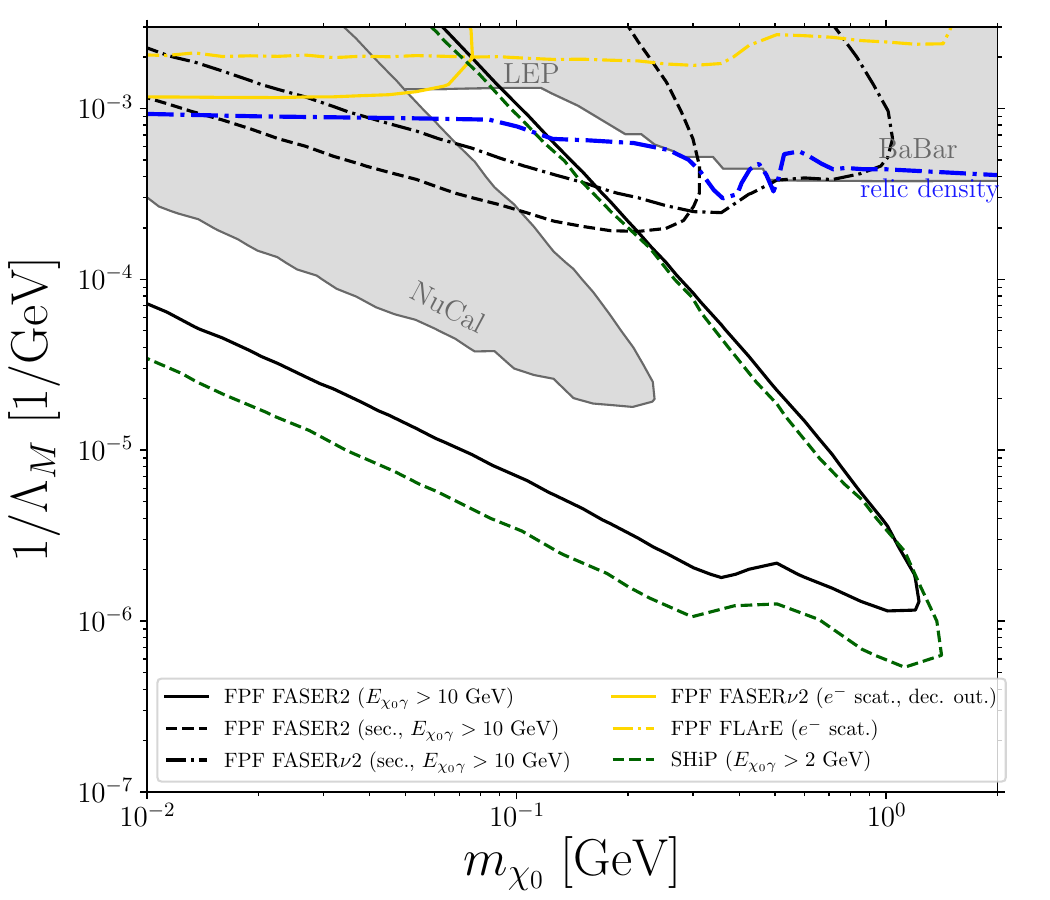}
  \caption{
    Sensitivity of FASER2 to inelastic MDM for $\Delta=10^{-3}$ (top) and $\Delta=0.05$ (bottom). 
    We consider two setups of the FASER2 experiment: the baseline scenario \textit{(left)} and the Forward Physics Facility containing additional detectors, \eg, FLArE \textit{(right)}. Note the lower energy threshold ($E_{\chi_0 \gamma}>1\,\gev$) for the photon coming from $\chi_1$ decay for the case $\Delta=10^{-3}$, which is needed due to kinematics of the process, to obtain a sufficient number of events. 
    For both benchmarks, we reproduce results of \cite{Dienes:2023uve} for primary LLP production (black and red solid lines for FASER and green dashed line for SHiP).
    Sensitivity reaches using secondary production (dashed and dash-dotted black lines) allow to cover the smaller lifetime regime, $d_{\chi_1}\sim 1\,\m$, which for $\Delta=0.05$ largly overlaps with the thermal relic density line, $\Omega_{\chi_0}h^2=0.1$.
    The electron scattering signature at FASER$\nu$2 (solid gold line) and FLArE (dot-dashed gold line) leads to sensitivities analogous to the case of elastic coupling studied in \cite{Kling:2022ykt}.
    }
  \label{fig:MDM}
\end{figure*}

\begin{figure*}[tb]
  \centering
  \includegraphics[width=0.49\textwidth]{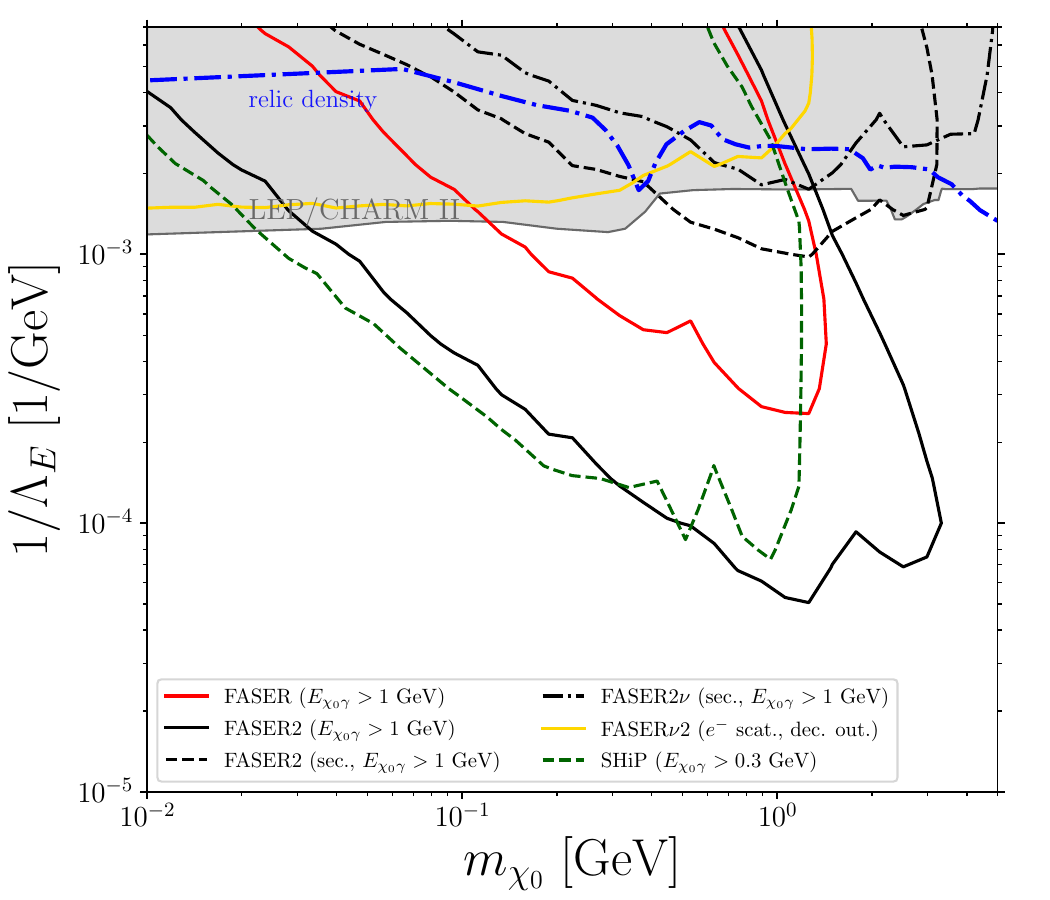}\hspace*{0.4cm}
  \includegraphics[width=0.49\textwidth]{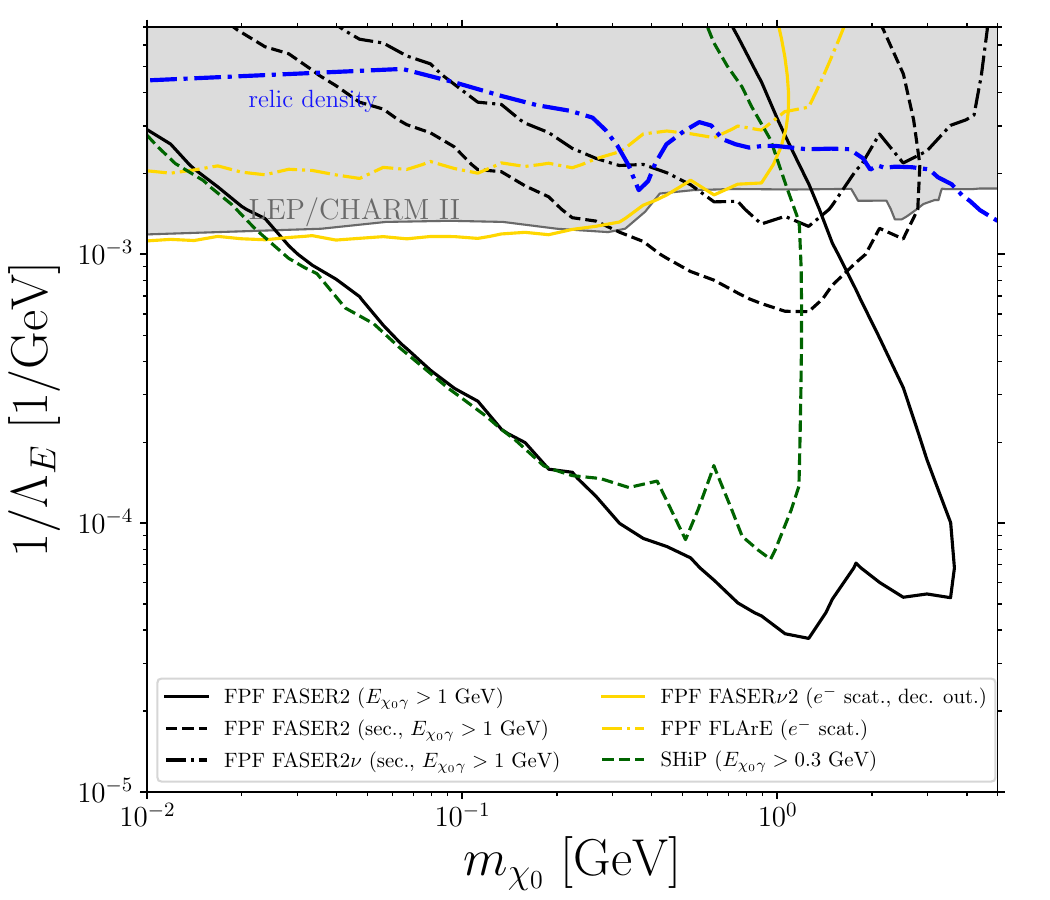}\vspace*{0.2cm}
  \includegraphics[width=0.49\textwidth]{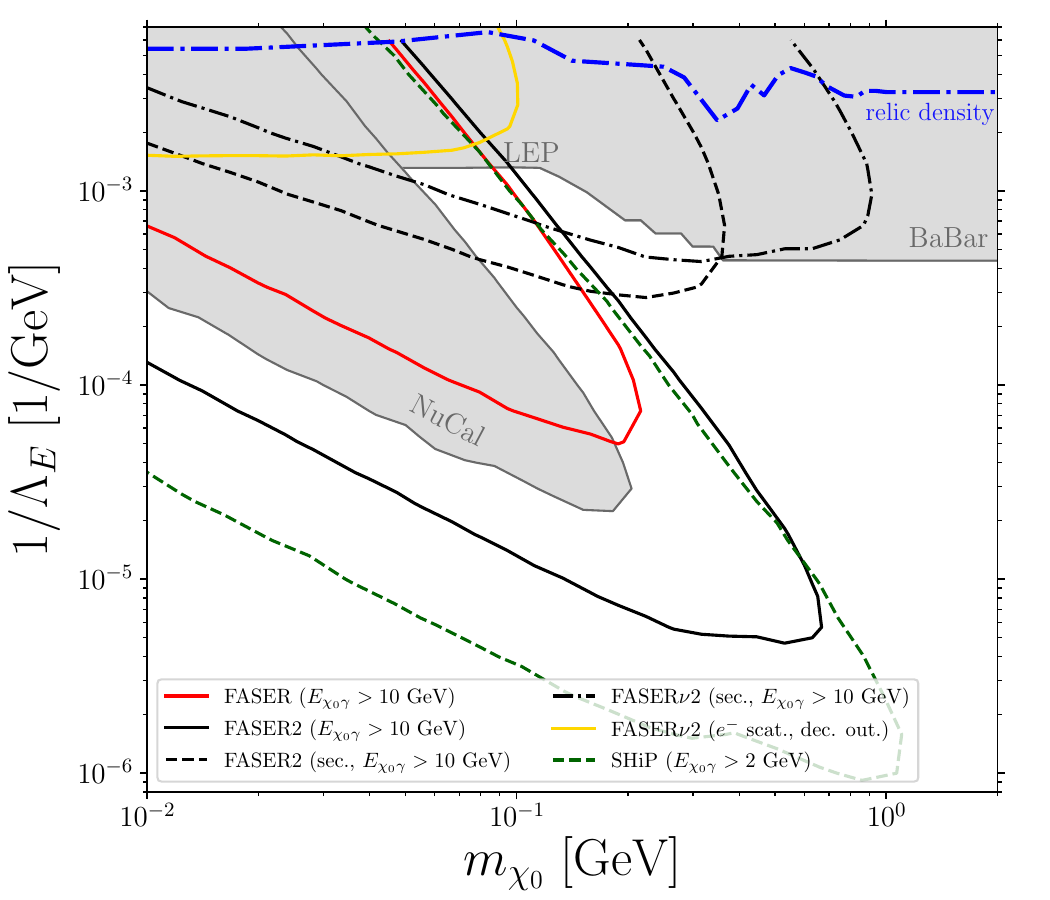}\hspace*{0.4cm}
  \includegraphics[width=0.49\textwidth]{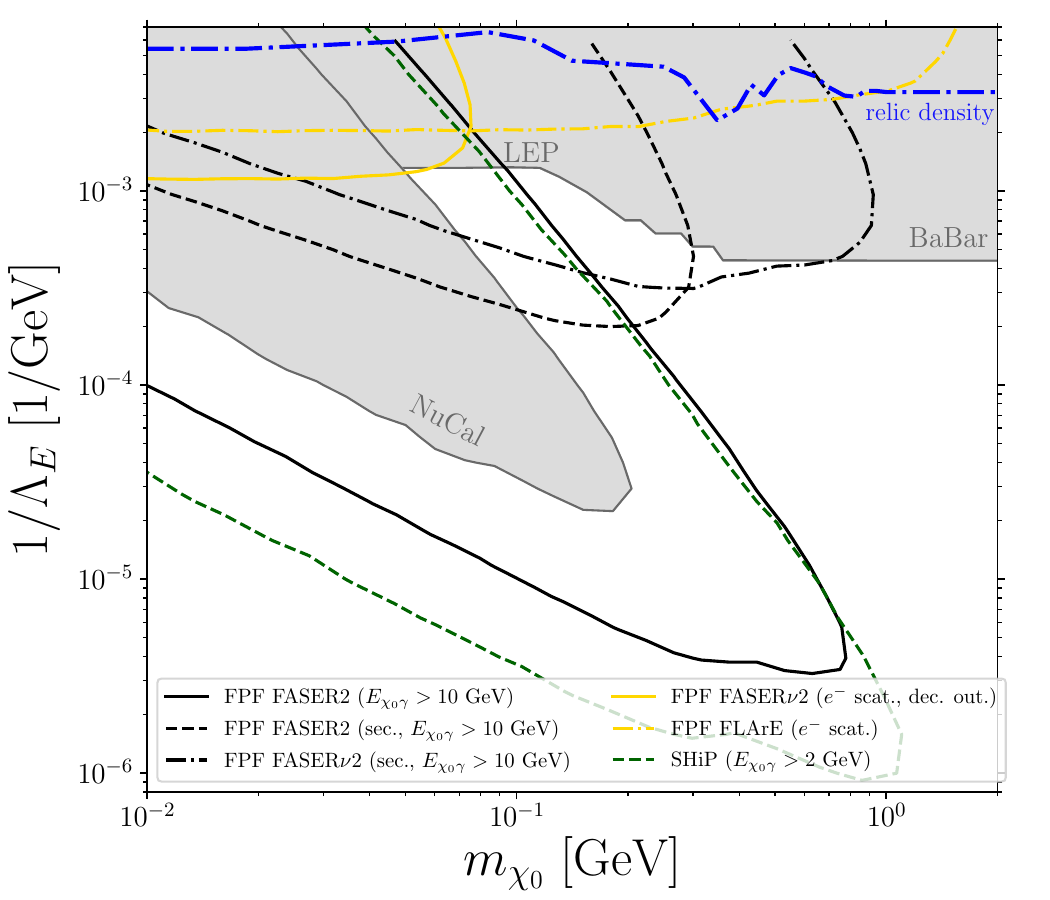}
  \caption{
    Same as \cref{fig:MDM} but for the EDM iDM.
    }
  \label{fig:EDM}
\end{figure*}

\paragraph{Secondary LLP production}
As described above, searches using displaced LLP decays at remote detectors are limited to sufficiently long-lived BSM species.
In addition to the main decay vessel of FASER, which is designed to study LLP decays, a specialized neutrino detector called FASER$\nu$ \cite{FASER:2019dxq,FASER:2020gpr} has been positioned in front of it.
It is an emulsion detector made of tungsten layers, which allowed the FASER collaboration to detect collider neutrinos for the first time \cite{FASER:2023zcr}.

Moreover, FASER$\nu$ can be used as a target to produce secondary LLPs in BSM scenarios with at least two DS species with different masses, which are produced in pairs in the forward direction of the detector.\footnote{The mechanism was illustrated in Fig. 1 of \cite{Jodlowski:2019ycu}.}
This process occurs through the upscattering of a lighter DS species into the heavier one, which acts as a LLP.
It was shown in \cite{Jodlowski:2019ycu,Jodlowski:2020vhr,Jodlowski:2023sbi} that such upscattering is particularly efficient in the coherent regime, involving interactions with the entire nucleus (\eg, tungsten for FASER$\nu$).
After the secondary LLP production, in our case $\chi_1$, it can decay within either the FASER$\nu$ or FASER detectors.\footnote{Different energy thresholds are set for each. We use the cuts shown in Tab. 1 of \cite{Jodlowski:2023sbi}, except for FASER detectors, where we use the energy thresholds used in the previous study dedicated to EDM/MDM iDM - Ref. \cite{Dienes:2023uve}.}

Thus, in this way, FASER may cover part of the $d \sim L \simeq 1\,\m$ region of the parameter space that is inaccessible by primary production. Moreover, such a regime is particularly interesting, \eg, it typically corresponds to the correct thermal relic density region - see Fig. 6 in \cite{Jodlowski:2019ycu} for the case of iDM coupled to a dark photon, which connects to the SM via kinetic mixing.
On the other hand, due to the form of the formula describing the secondary production cross section, see \cref{eq:Prim_iDM}, the number of decays of LLPs produced in this way has an additional dependence on the coupling constant squared.
Therefore, the secondary production of LLP can only be efficient for sufficiently large values of the coupling constant.

Secondary $\chi_1$ production is dominated by coherent scattering with an entire nucleus due to the $Z^2$ enhancement, compared to only $Z$-dependent incoherent scatterings.
We obtained the following integrated cross sections for such processes:
\be
  \text{EDM/MDM:}  \
   \sigma_{\chi_0 N \to  \chi_1 N} \simeq & \frac{4 \alpha_{\mathrm{EM}} Z^2}{\Lambda_{\mathrm{E/M}}^2} \times \\
  &\left(\log \left(\frac{d}{1/a^2 - t_{max}}\right)-2\right), \\
  \text{AM/CR:}  \
   \sigma_{\chi_0 N \to  \chi_1 N} &\simeq (a_\chi^2,b_\chi^2) \times(\alpha_{\mathrm{EM}}\, Z^2 \,d), 
  \label{eq:Prim_iDM}
\ee
where $a=111 Z^{-1/3}/m_e$, $d=0.164\, \gev^2 A^{-2/3}$, $m_e$ is the electron mass, $Z$ ($A$) is the atomic number (mass number) of a nucleus, and $t_{\mathrm{max}} \simeq -(m_{\chi_1}^2-m_{\chi_0}^2)^2/(4E_1^2)$. We used the form factor given in Eq. B2 in \cite{Jodlowski:2023sbi}.
In addition, to obtain \cref{eq:Prim_iDM}, we used the method described in \cite{Dusaev:2020gxi,Jodlowski:2023sbi}. The full derivation, including the square of the amplitude, can be found in the accompanying Mathematica notebook.

\paragraph{Electron scattering signature}
In the case of elastic interactions described by \cref{eq:lagr}, the DS species is stable, and the main signature at the intensity frontier experiments is elastic scattering with electrons \cite{Chu:2020ysb,Kling:2022ykt}.
Following these works, we consider this signature for the iDM scenario.

The integrated cross sections are given in \cref{eq:Prim_e_integr}, while \cref{eq:Prim_e_dsigmadER} gives their differential form.
The latter is needed to impose the energy- and angle-dependent cuts for the FASER$\nu$2 and FLArE electron scattering searches \cite{Batell:2021blf,Jodlowski:2020vhr,Kling:2022ykt}.
The electron recoil angle is measured relative to the beam direction, and it reads
\be
  \cos\theta_e= \frac{(E_{\chi_0}+m_e)E_R+(m_{\chi_1}^2-m_{\chi_0}^2)/2} {\sqrt{E_{\chi_0}^2-m_{\chi_0}^2}\sqrt{E_R(E_R+2m_e)}}
\,.
\ee
We apply cuts from \cite{Batell:2021blf} - Tab. 1 and 2 therein; also see Tab. 1 in \cite{Jodlowski:2023sbi}.
Furthermore, following \cite{Jodlowski:2020vhr}, we require that the $\chi_1$ produced in the upscattering with electrons inside the FASER$\nu$ decays outside of it and the main FASER decay vessel.
In \cref{fig:Prim_e}, we show the mass dependence of the cross section for upscattering ($\chi_0 T \to \chi_1 T$, $T=\{N,e^-\}$) on nucleus (left) and on electron (right) for each of the iDM operator we consider. We consider two energies of the incoming $\chi_0$, $E_1=100\,\gev$ and $E_1=1000\,\gev$.
We assumed a coupling constant for each model equal to $1$ in units of $\gev$.

Since the magnitude of the upscattering cross section plays a crucial role in the significant secondary production, we also present the cross section for the Primakoff production of ALP \cite{Dusaev:2020gxi}.
For this model, the Primakoff process actually plays the role of primary LLP production, since it dominates the contributions of meson decays.
However, for the extended ALP scenario with its coupling to a photon and a dark photon, called the dark axion portal \cite{Kaneta:2016wvf,Ejlli:2016asd}, the analogous Primakoff process with the on-shell photon replaced by a dark photon has the same form, except it is smaller by a factor of $3/2$ \cite{Jodlowski:2023sbi}. 
As a result, such secondary production allows FASER2 to also cover part of the $d \sim 1\,\m$ region of the parameter space of this model, see Fig. 2-4 in \cite{Jodlowski:2023sbi}.

In the iDM case, the character of the interactions is clearly visible - the dimension-5 operators lead to sizable cross sections, while results for operators of dimension-6 are suppressed. 
Therefore, one can expect to extend the sensitivity reach of FASER2 using displaced LLP decays thanks to secondary production only for dimension-5 operators.
On the other hand, the cross sections given by \cref{eq:Prim_iDM} are enhanced by a factor of $Z^2$, hence the number of such events may exceed the number of electron scattering, which is only proportional to $Z$.
Thus, it is worthwhile to investigate the upscattering process acting as either secondary LLP production or scattering with electrons for all iDM scenarios.

The origin of the suppression for AM/CR is similar to the reason why $\chi_1 \to \chi_0 e^+ e^-$ decay is possible, while $\chi_1 \to \chi_0 \gamma$ is not.\footnote{In fact, the squared amplitude of the $\chi_1 \to \chi_0 e^+ e^-$ and $\chi_0 e^- \to \chi_1 e^-$ processes are related to each other by crossing symmetry.}
The $\chi_0{\text -}\chi_1{\text -}\gamma$ coupling for the AM/CR operator depends on the four-momentum of the photon.
Hence, the square of the amplitude vanishes for on-shell photon, while for off-shell photon, which mediates either the upscattering or decay of $\chi_1$, its propagator cancels this common factor of the coupling, rendering the square of the amplitude non-zero.
This behavior also illustrates why the sensitivity reaches for electron scattering for elastic DS with EM form factors for EDM/MDM are much stronger than for the AM/CR operator \cite{Kling:2022ykt,Chu:2020ysb}.

\subsection{Simulation of LLP signatures \label{sec:experiments}}

We implement the signatures of the iDM operators introduced in previous subsection within modified $\tt FORESEE$ \cite{Kling:2021fwx} package.
The implementation includes, \eg, the production of LLPs as described in paragraph a) above.
In particular, we used the SM meson spectra generated by $\tt EPOSLHC$ \cite{Pierog:2013ria} and $\tt Pythia$ \cite{Sjostrand:2014zea}. We then use the routines of the modified $\tt FORESEE$ to obtain the $\chi_0$-$\chi_1$ yield, and the number of events corresponding to each signature discussed in \cref{sec:iDM_production}.

We explore the prospects of iDM in several intensity frontier experiments - see Tab. 1 in \cite{Jodlowski:2023sbi} for a specification of the detector parameters used in our simulation and the discussion that follows there.
In particular, we consider the LHC far-forward detectors such as FASER \cite{Feng:2017uoz,Feng:2017vli,FASER:2022hcn}, FASER2 \cite{FASER:2018ceo,FASER:2018bac,FASER:2021ljd}, and FASER$\nu$2 \cite{FASER:2019dxq,FASER:2020gpr,Batell:2021blf,Anchordoqui:2021ghd}.
We also study prospects of the proposed MATHUSLA \cite{Chou:2016lxi,Curtin:2018mvb} detector and the Forward Physics Facility (FPF) \cite{MammenAbraham:2020hex,Anchordoqui:2021ghd,Feng:2022inv}. 
The latter is a proposed facility housing not only an alternative, larger versions of FASER2 and FASER$\nu$2, but also another detectors searching for feebly-interacting BSM particles, \eg, AdvSND \cite{Boyarsky:2021moj}, FLArE \cite{Batell:2021blf}, and FORMOSA \cite{Foroughi-Abari:2020qar}.
We also study limits from previous beam dump experiments, such as CHARM \cite{CHARM:1985anb} and NuCal \cite{Blumlein:1990ay,Blumlein:2011mv}, as well as the proposed SHiP detector \cite{SHiP:2015vad,Alekhin:2015byh}.

We note that the experiments described above have been investigated for EDM/MDM iDM in \cite{Dienes:2023uve}, and we follow their discussion of simulating the LLP decays.
In particular, we adapt their energy thresholds for all of these experiments.
In addition, we examine other iDM signatures in the long-lived regime, which were described in the previous subsection.

Our choice of detectors is motivated by the desire to cover as much parameter space as possible.
The selected experiments differ from each other in key physical characteristics, such as, \eg, the distance $L$, luminosity and energy of the proton beam, size, and visible energy threshold.
As a result, as shown in the next section, they allow coverage of a variety of iDM scenarios, similar to many popular LLP scenarios, such as the renormalizable portals or an ALP coupled to a pair of photons; see extensive discussion in recent reviews \cite{Battaglieri:2017aum,Beacham:2019nyx,Alimena:2019zri}.

\section{Results\label{sec:results}}

In this section, we describe the prospects of detecting iDM with EM form factors in beam dump experiments and LHC far-forward detectors.
All models described in \cref{sec:models} are characterized by three free parameters: ($\Lambda$, $m_{\chi_0}$, $\Delta$) - the coupling constant, the mass of the stable iDM species, and the mass splitting.

As discussed in \cref{sec:relic_density}, mass splittings $\Delta$ plays a key role in determining the iDM relic density. 
Therefore, we present our results in the ($m_{\chi_0}$, $\Lambda$) plane for several values of $\Delta$.
In the case of EDM/MDM, we study the two benchmarks from \cite{Dienes:2023uve}, corresponding to $\Delta=10^{-3}$ and $\Delta=0.05$, respectively. 
On the other hand, for AM/CR, we investigate larger values of $\Delta$: $0.05$, $0.2$, and $1$.

This choice is motivated by the intention to examine the regime of long-lived $\chi_1$, which for dimension-6 operators, is shifted towards larger values of the mass splittings, cf. \cref{eq:Gamma_EDM_MDM} and \cref{eq:Gamma_AM_CR}.
We also note that for all iDM scenarios, our simulation is adapted to any $\Delta$, and can be easily modified by, \eg, changing the experimental cuts.

\subsection{Sensitivity reach for dimension-5 operators}
\label{sec:models_dim5}

In \cref{fig:EDM,fig:MDM}, we present the results for MDM and EDM iDM, respectively.
For the displaced LLP decays produced by vector meson decays, which is the main iDM signature in the long-lived regime, we reproduce the results from \cite{Dienes:2023uve}. The legend of each plot includes the visible energy threshold imposed on the $\chi_1$ decays in each experiment. The constraints shown were derived by LEP \cite{L3:2003yon,Fortin:2011hv} and CHARM-II \cite{CHARM:1983ayi}, while we also show the bounds from NuCal.
In particular, note that the benchmark used in the first line of \cref{eq:dbar} is in the upper right area of the FASER2 sensitivity plot for both iDM operators, which agrees with our discussion following this equation.

For dimension-5 operators, our main results are the sensitivity lines obtained from the secondary LLP production.
As described in \cref{sec:iDM_production}, this process proceeds through coherent upscattering, $\chi_0 N \to \chi_1 N$, on the nuclei of the tungsten layers of the FASER$\nu2$ emulsion detector located in front of the main decay vessel, FASER2.
Since the distance between the two detectors is $\sim O(1)\,\m$, secondary production allows to 
cover such short-lived LLP regime, which is otherwise challenging to probe because of the exponential suppression, see \cref{eq:p_prim}.
We note that this configuration of the FASER detectors presents advantages over a typical beam dump experiment, in which there is a large separation, $\sim O(100)\,\m$, between the LLP production and detection points.

After $\chi_1$ is produced by upscattering, it can decay in FASER$\nu2$ (dashed-dotted black lines) or inside FASER2 (dashed black lines). 
As can be clearly seen, this mode of $\chi_1$ production allows covering the region of larger mass and coupling values than the primary production.
This is particularly important for MDM, where it can help cover part of the parameter space where $\chi_0$ is a thermal DM candidate.

We also present results for the electron scattering signature at FASER$\nu$2 (gold solid line) and FLArE (gold dot-dashed line).
It covers the low-mass regime that complements the sensitivity reach obtained using decays of $\chi_1$ produced in either primary or secondary production processes.
The projection lines derived in such a way are generally weaker than those using secondary LLP production.
This originates from the scaling behavior of the number of events with the atomic number, which is $Z$ for the electron scattering and $Z^2$ for the upscattering production.
We also note that the FASER$\nu$2 sensitivity ends at $m_{\chi_0} \sim 0.1\,\gev$ due to the imposed condition on $\chi_1$ decays - they must take place outside both FASER detectors - which is best illustrated in the bottom panels of \cref{fig:EDM}.

In summary, secondary $\chi_1$ production allows to cover a part of the short-lived regime of EDM/MDM that is inaccessible by primary LLP production.
Depending on the value of the mass splitting $\Delta$, it gives comparable or stronger bounds than the scattering signature, which for iDM gives similar results to the elastic DS with EM form factors scenario \cite{Kling:2022ykt,Chu:2020ysb}.
As a result, a part of the allowed region of the MDM parameter space that corresponds to a successful thermal freeze-out of $\chi_0$ will be covered by FASER2 during the high luminosity era of the LHC.

\begin{figure*}[tb]
  \centering
  \includegraphics[width=0.45\textwidth]{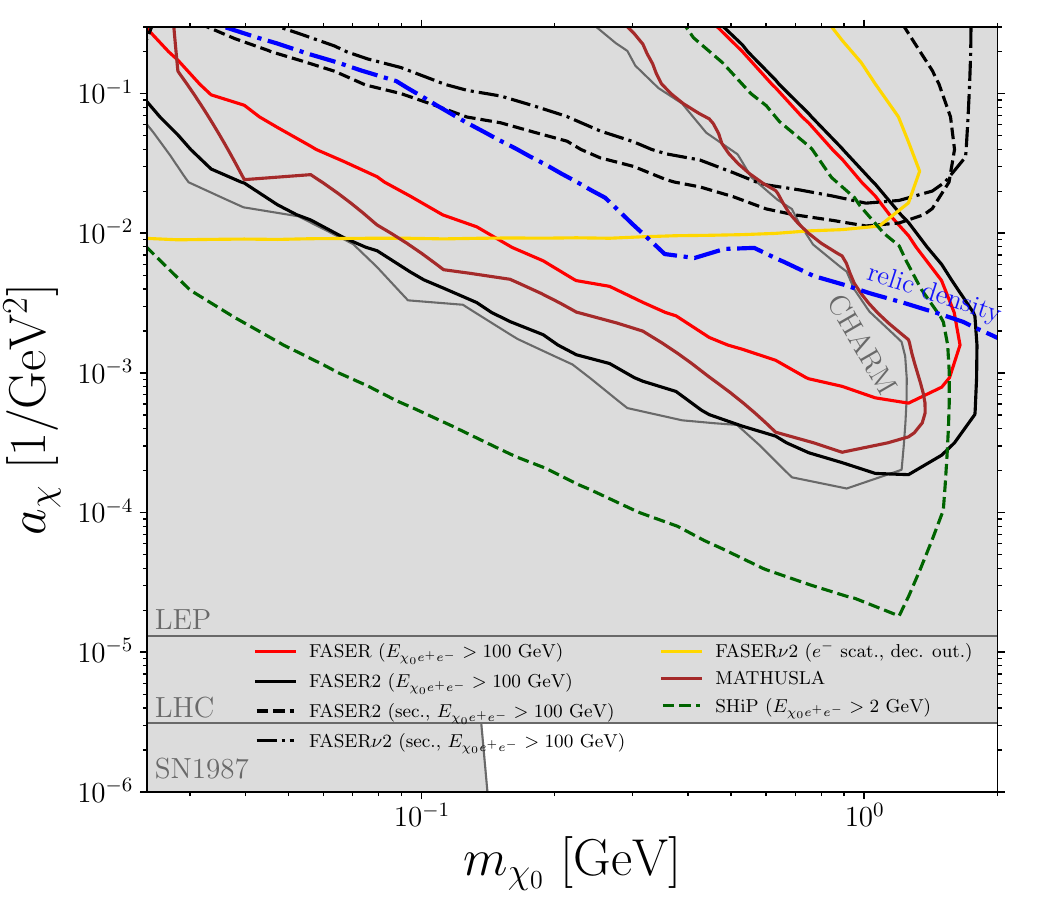}\hspace*{0.4cm}
  \includegraphics[width=0.45\textwidth]{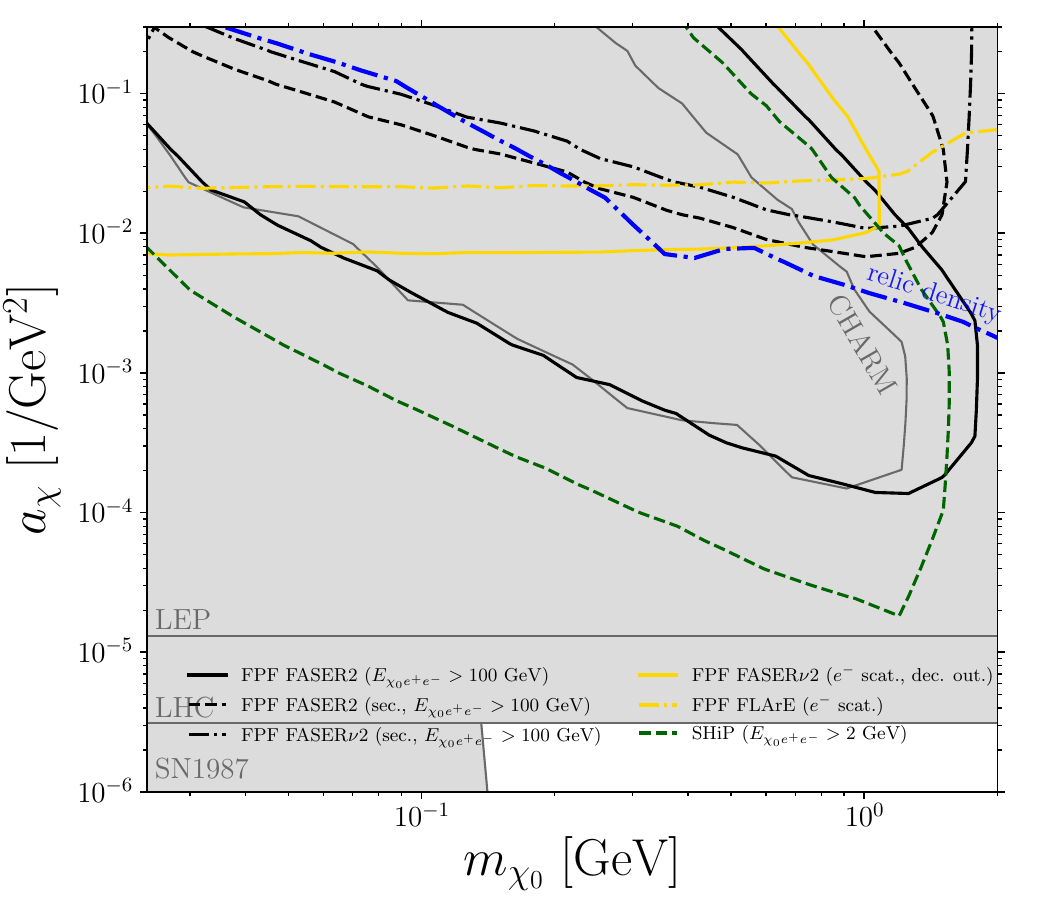}\vspace*{0.25cm}
  \includegraphics[width=0.45\textwidth]{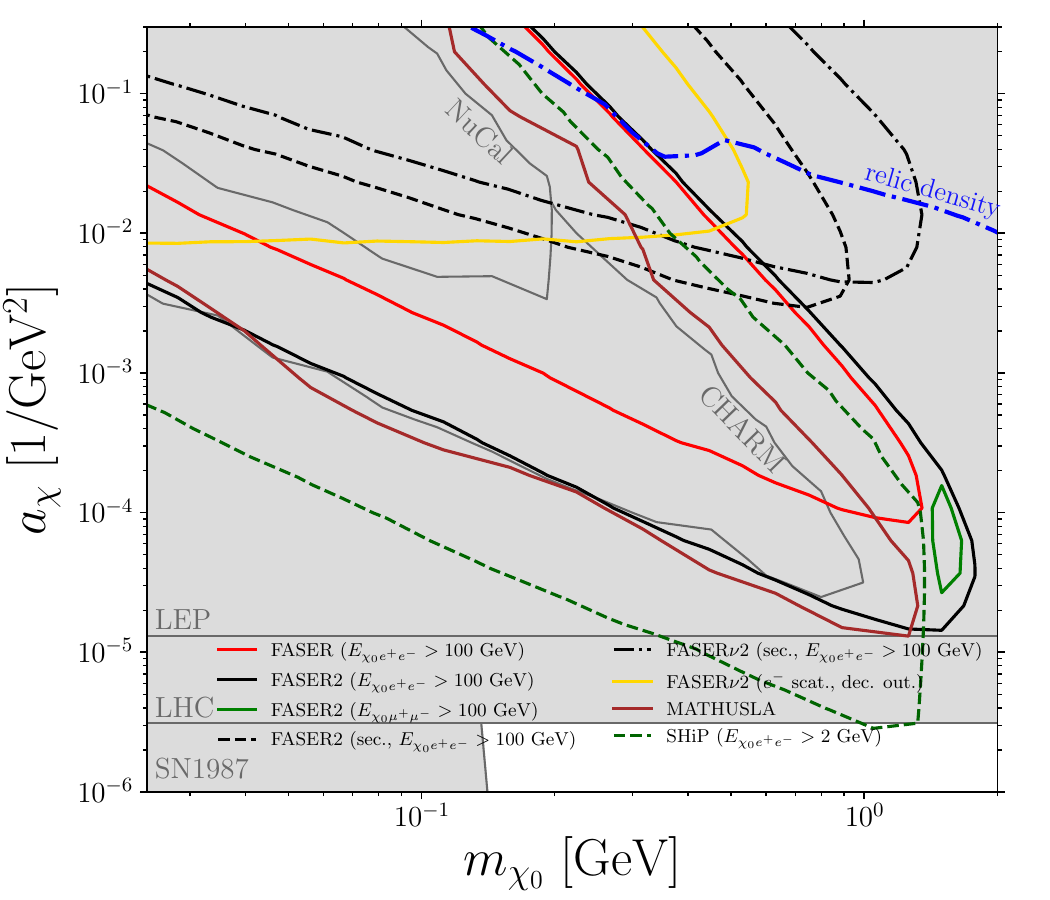}\hspace*{0.4cm}
  \includegraphics[width=0.45\textwidth]{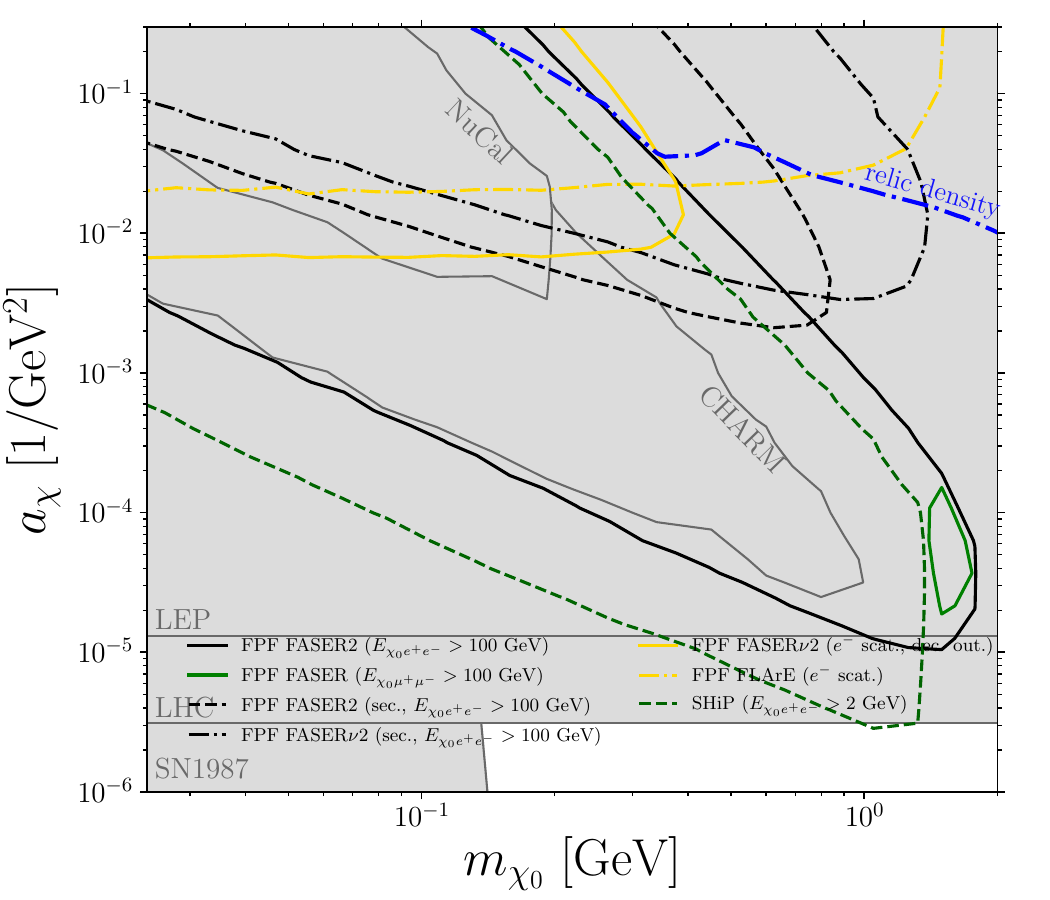}\vspace*{0.25cm}
  \includegraphics[width=0.45\textwidth]{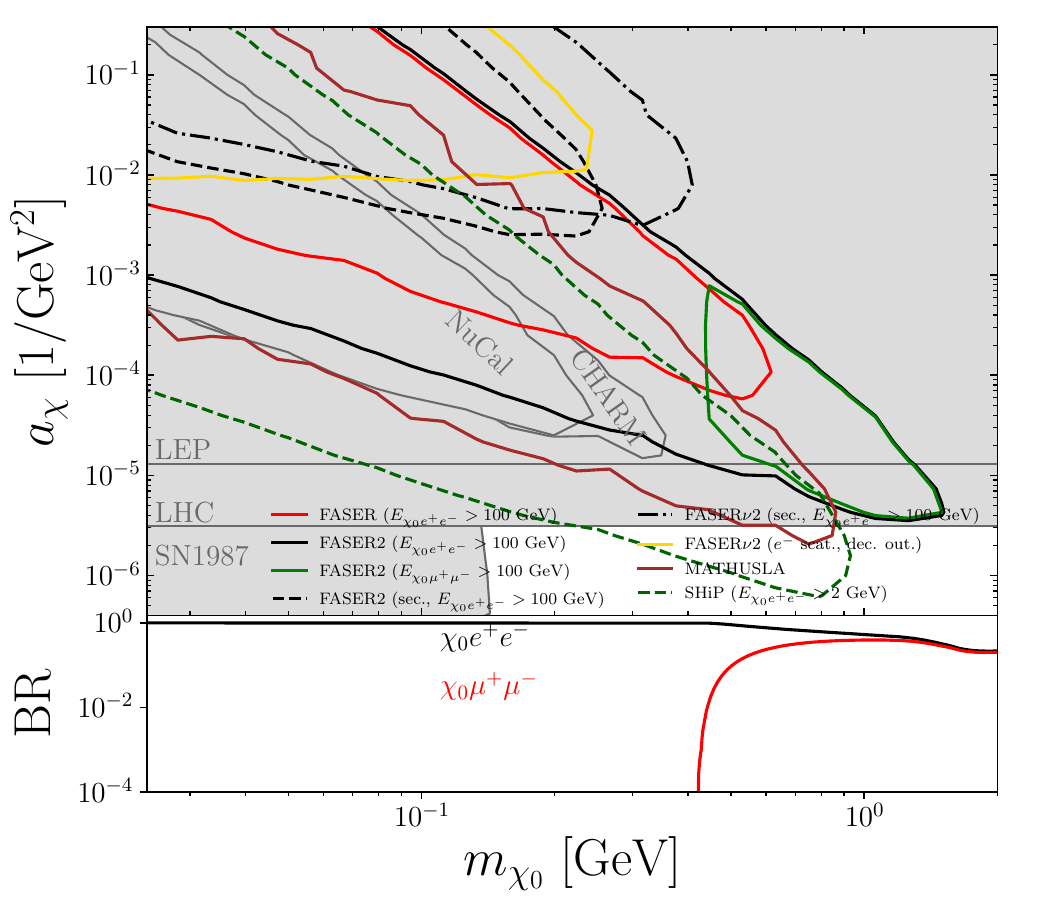}\hspace*{0.4cm}
  \includegraphics[width=0.45\textwidth]{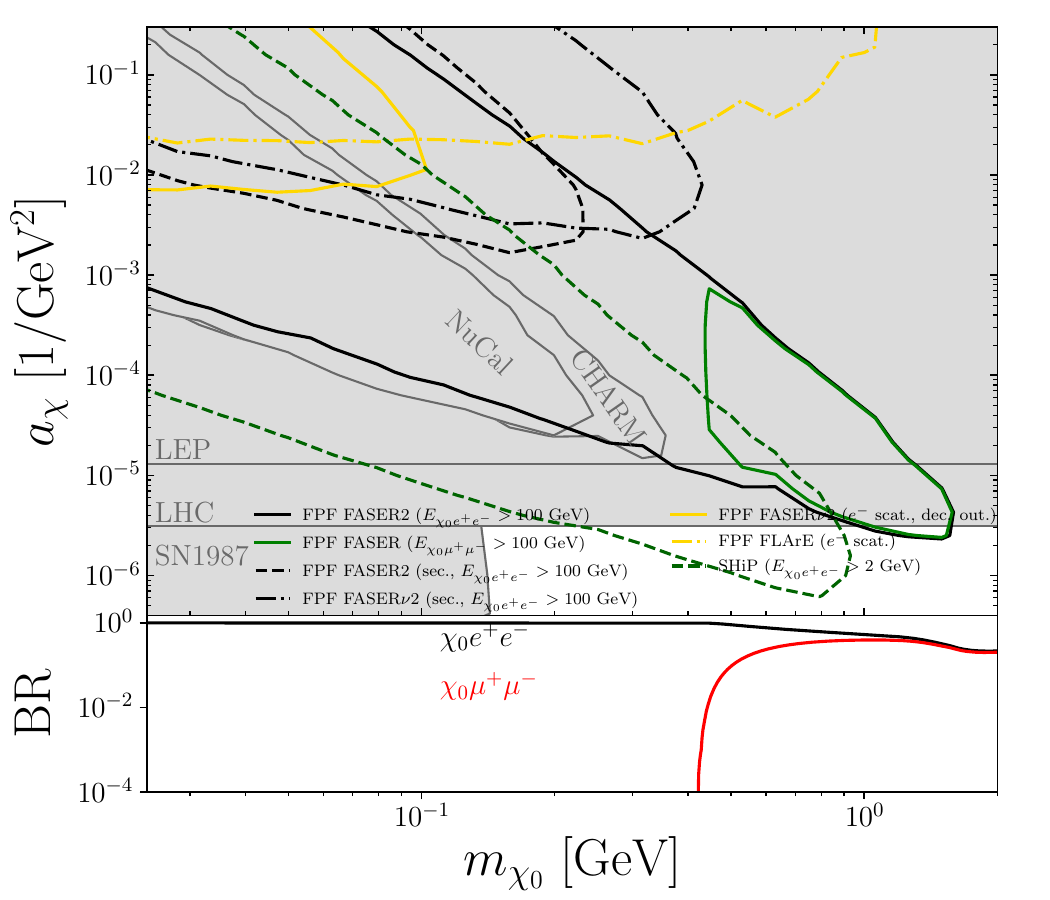}
  \caption{
    Sensitivity of FASER2 to AM iDM for $\Delta=0.05$ (top), $\Delta=0.2$ (middle), and $\Delta=1$ (bottom). Note that the line $\Omega_{\chi_0}h^2=0.1$ shifts significantly upwards with increasing $\Delta$ and does not actually exist for $\Delta=1$ - see discussion in \cref{sec:relic_density}.
    For large values of $m_{\chi_1}$ and $\Delta$, $\chi_1 \to \chi_0 \mu^+ \mu^-$ decays are possible - we show the branching ratios for the particularly relevant case of $\Delta=1$.
    We used \cref{eq:Gamma_chi1_tot} and the decay modes not shown are hadronic.
    See caption under \cref{fig:MDM} for details on the legend of the presented sensitivity reaches.
  }
  \label{fig:AM}
\end{figure*}

\begin{figure*}[tb]
  \centering
  \includegraphics[width=0.45\textwidth]{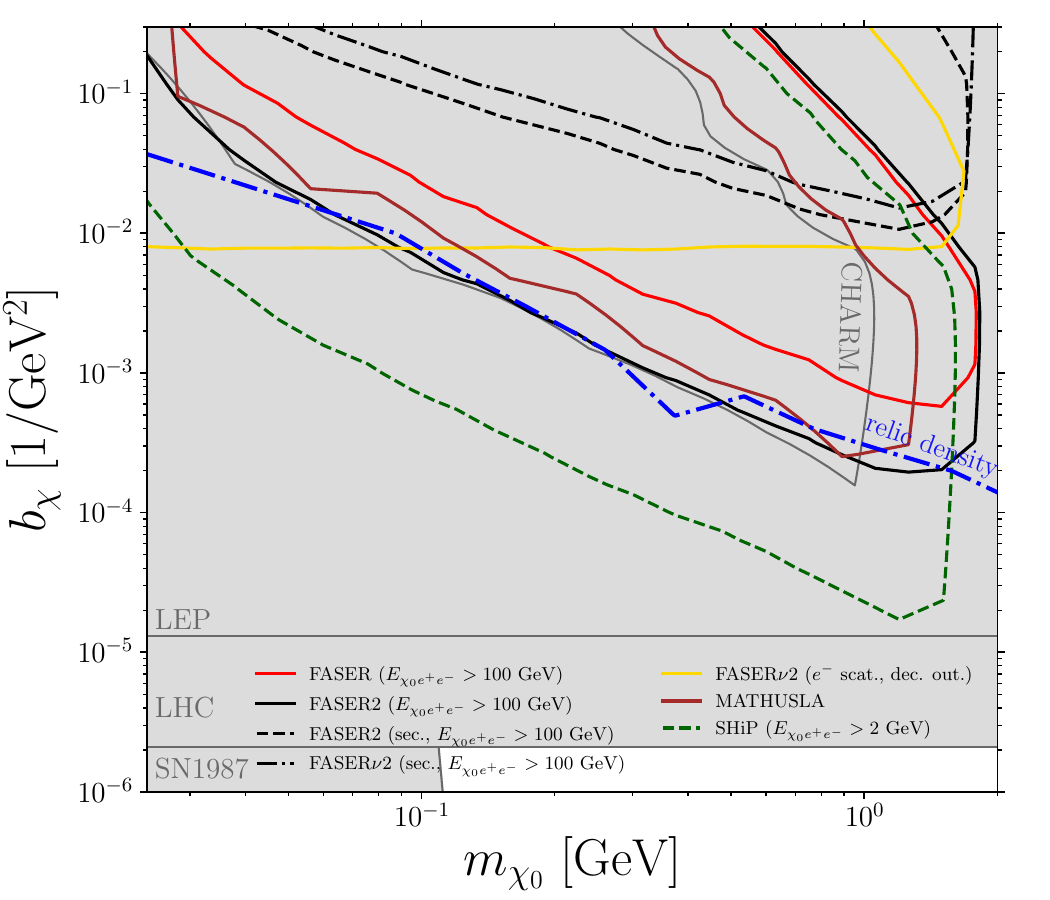}\hspace*{0.4cm}
  \includegraphics[width=0.45\textwidth]{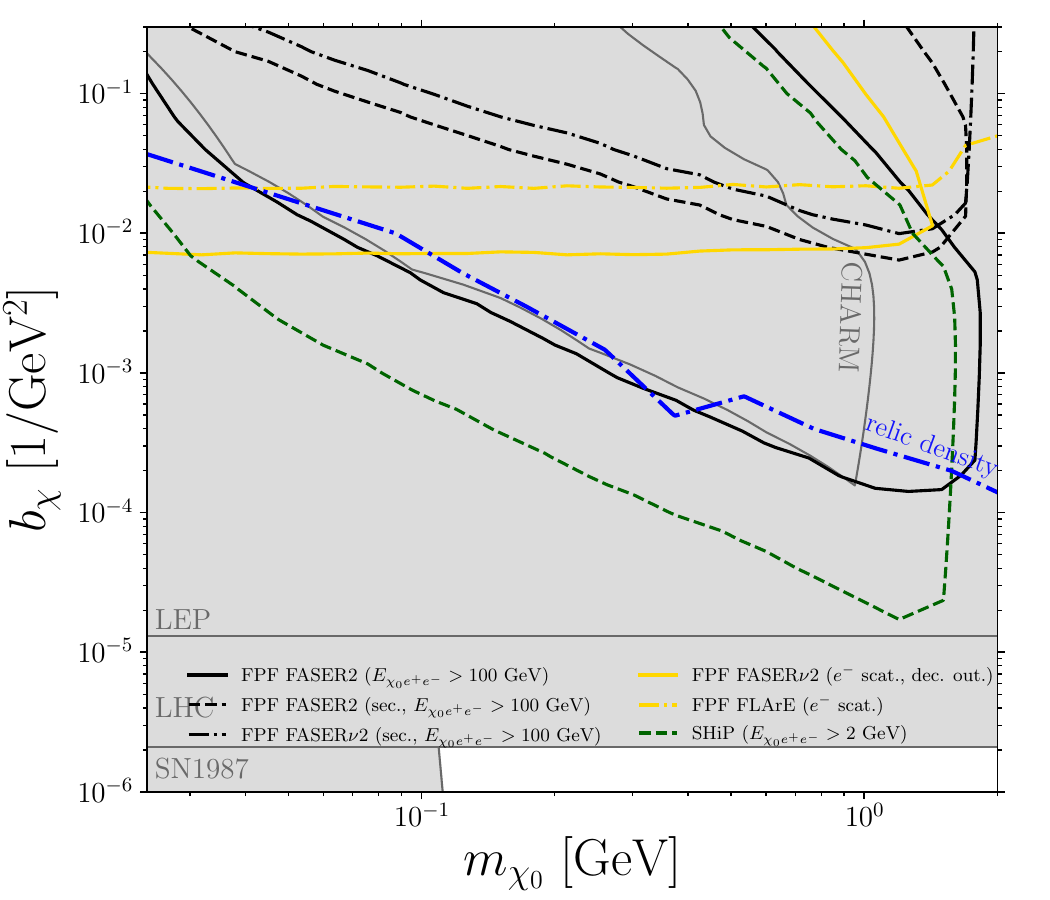}\vspace*{0.25cm}
  \includegraphics[width=0.45\textwidth]{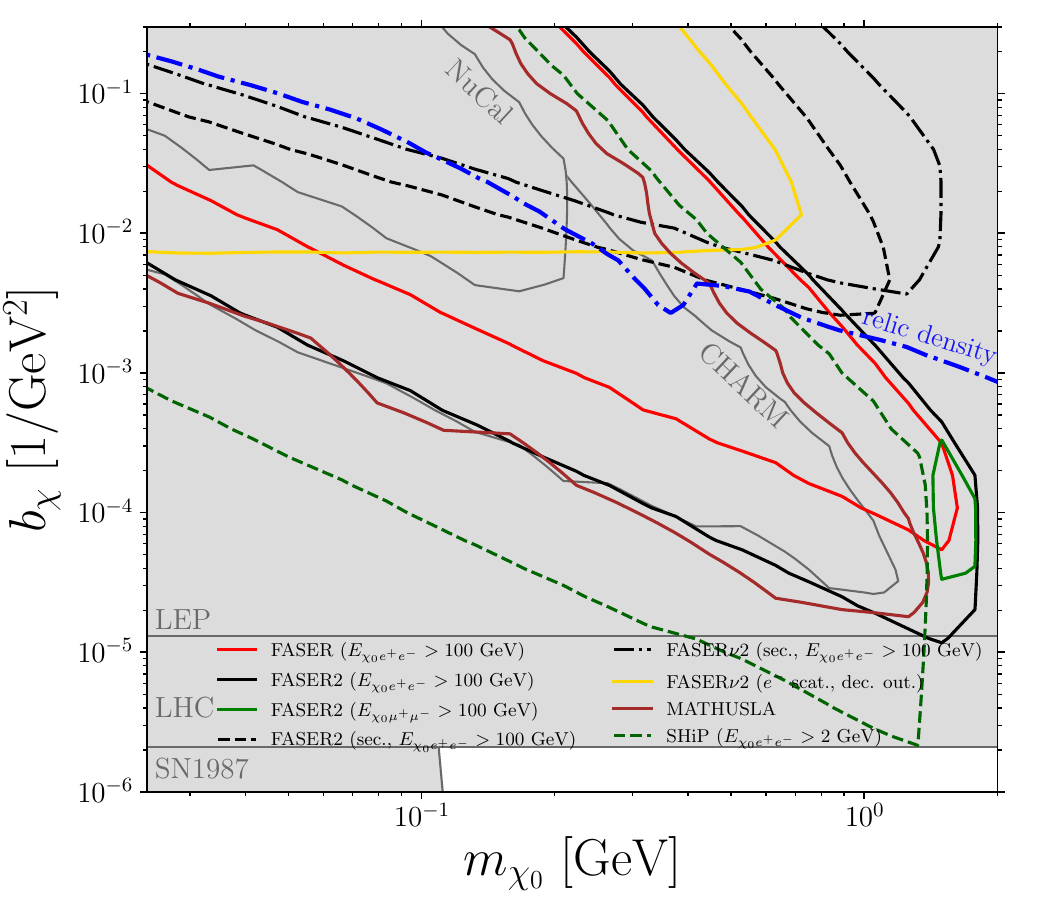}\hspace*{0.4cm}
  \includegraphics[width=0.45\textwidth]{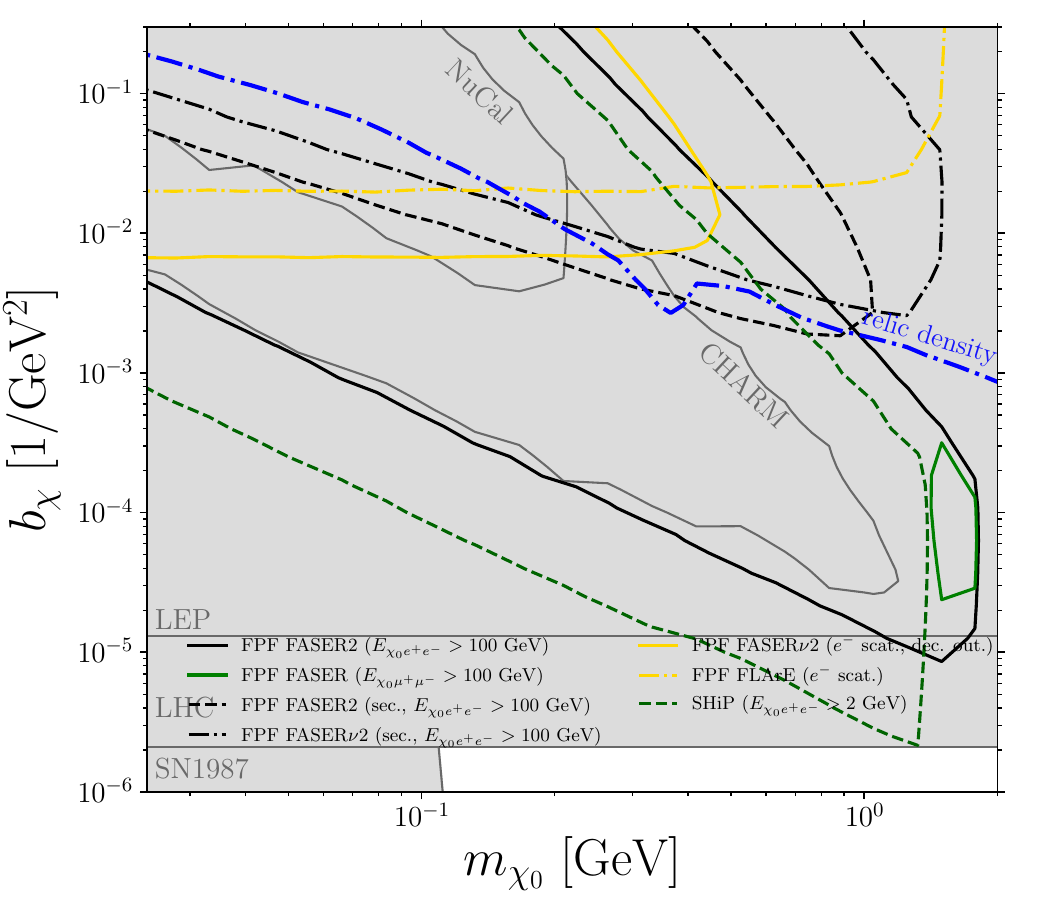}\vspace*{0.25cm}
  \includegraphics[width=0.45\textwidth]{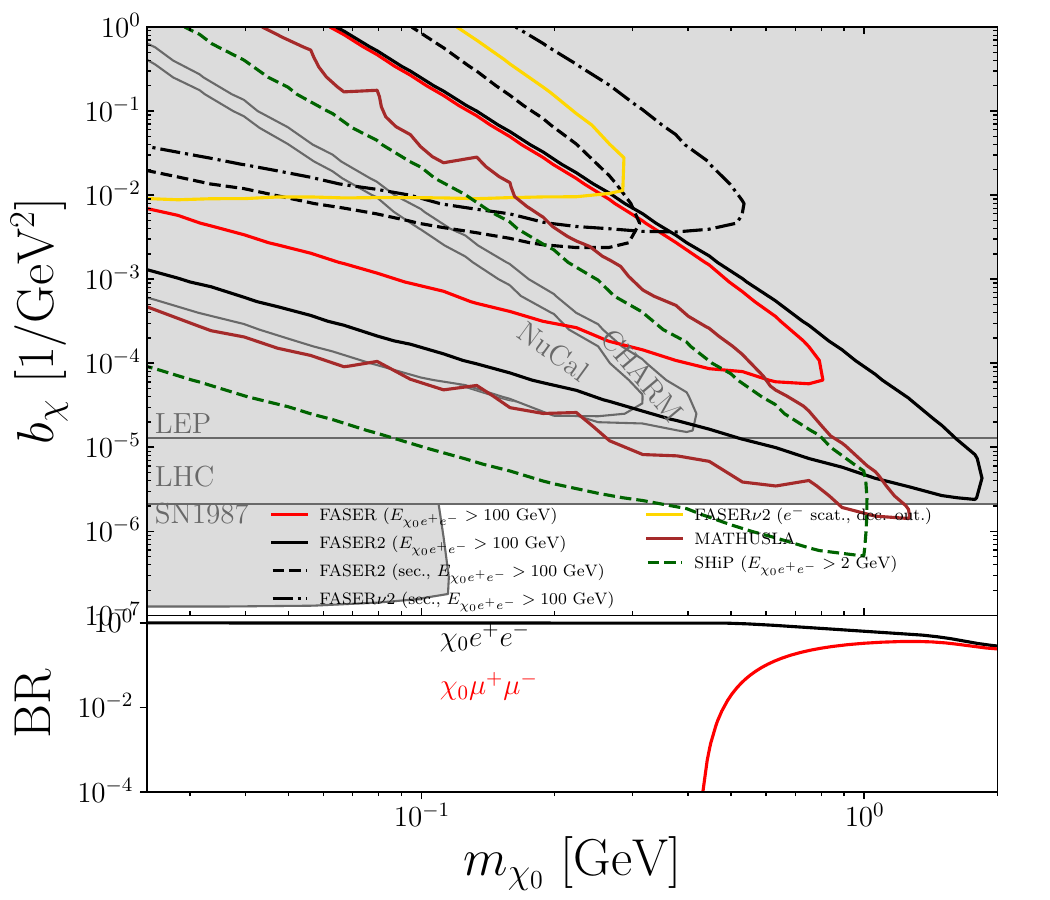}\hspace*{0.4cm}
  \includegraphics[width=0.45\textwidth]{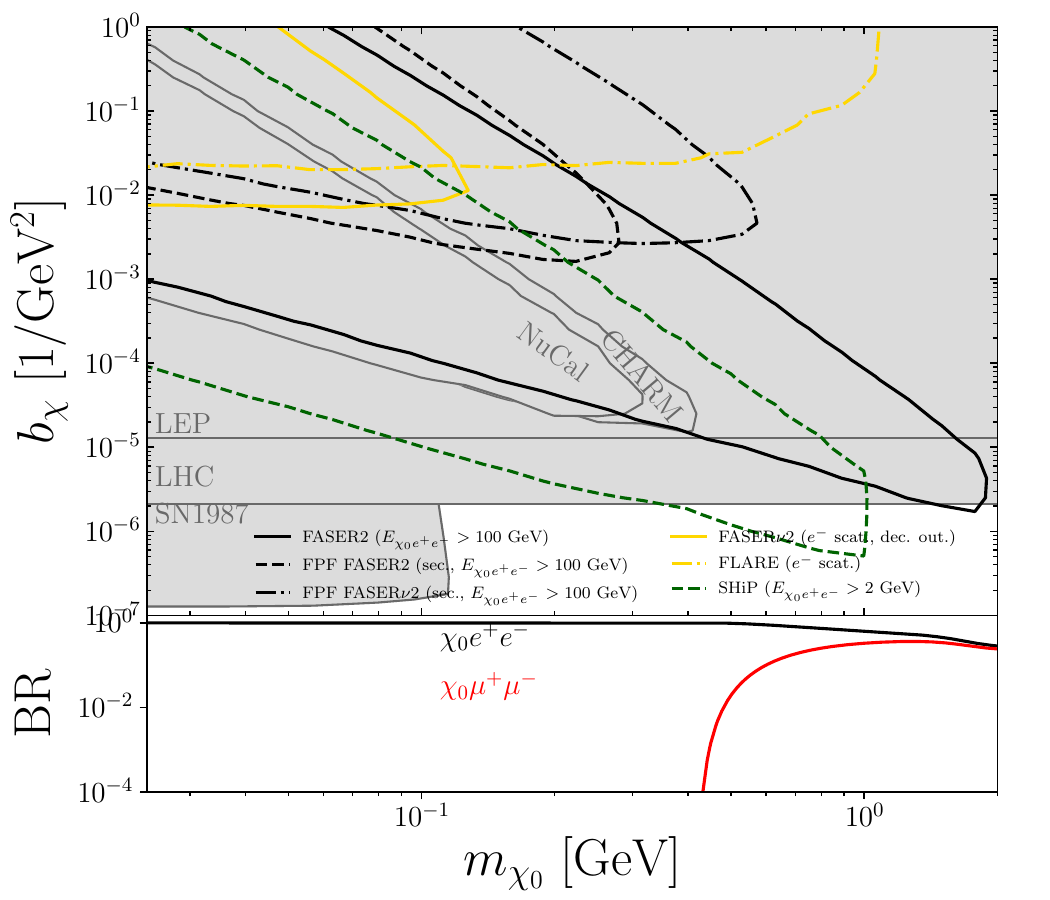}
  \caption{
    Same as \cref{fig:AM} but for the CR operator iDM.
    }
  \label{fig:CR}
\end{figure*}

\subsection{Sensitivity reach for dimension-6 operators}
\label{sec:models_dim5}

In \cref{fig:AM,fig:CR}, we show our results for AM and CR operator iDM. 
Both scenarios are mainly constrained by the bounds derived for the elastic coupling case, which were discussed in \cite{Kling:2022ykt,Chu:2020ysb}.
These bounds were obtained using the results of LEP \cite{L3:2003yon,Fortin:2011hv}, SN1987A \cite{Chu:2019rok}, and the LHC \cite{CMS:2017zts,Arina:2020mxo}. We also show the exclusion plots obtained by us using NuCal.
Moreover, similarly to the elastic case, in the AM/CR iDM scenario, the correct relic density of $\chi_0$ is obtained for the parameter space that is already excluded.

As discussed in \cref{sec:LLP_sign}, the upscattering cross section for AM/CR is suppressed by several orders of magnitude compared to the EDM/MDM iDM, see \cref{fig:Prim_e}.
As a result, the secondary $\chi_1$ production does not play a significant role, and the resulting sensitivity reach is comparable to the electron scattering signature, which in turn is similar to the scenario of elastic DS with EM form factors investigated in \cite{Kling:2022ykt}.

The main difference between that BSM scenario and ours is the possibility of $\chi_1$ decays.
As \cref{eq:Gamma_AM_CR} indicates, the leading decay width of $\chi_1$ is phase-space suppressed, therefore, the lifetime regime that can be probed by intensity frontier searches is moved to larger values of $\Delta$; see also discussion above \cref{eq:dbar}.
On the other hand, such large mass splittings make the co-annihilation rate insufficient for successful freeze-out of $\chi_0$, leading to its overabundance.

To resolve this tension between the intensity frontier and cosmological considerations, in \cref{fig:AM,fig:CR}, we show the results for three values of the mass splitting $\Delta$: $0.05$, $0.2$, and $1$.
In all cases, the displaced decays of $\chi_1$ will probe a significantly larger parameter space than the electron scattering signature, whose reach is again similar to the scenario of elastic DS with EM form factors studied in \cite{Kling:2022ykt}.

Such behavior can be explained by the fact that the LLP decays are not suppressed by the electron number density like the electron scattering signature. 
Moreover, similar to the case discussed in \cref{sec:models_dim5}, secondary production allows to cover a similar or larger area of parameter space than electron scattering.

On the other hand, for the largest value of $\Delta$ we consider, the displaced $\chi_1$ decays will allow FASER2, and in particular SHiP, to cover a part of the parameter space not covered by either the LHC or SN1987 (which is adapted from the elastic case studied in~\cite{Chu:2019rok}).
Although in such a regime co-annihilation is suppressed, and $\chi_1$ typically cannot freeze-out to the observed DM relic abundance, it is still interesting that intensity frontier experiments will be able to extend the sensitivity reach beyond the LEP and LHC bounds.

In summary, the long-lived nature of $\chi_1$ in the AM/CR scenario of iDM will allow LHC detectors such as FASER2 and MATHUSLA, and beam dumps like SHiP, to set significant exclusion limits that will be complementary to the existing bounds using other terrestrial searches. 
Moreover, in the regime of large mass splitting, such a signature will even allow going beyond such limits for a certain region of parameter space corresponding to $m_{\chi_0} \sim 1 \gev$.

\section{Conclusions\label{sec:conclusions}}

In this work, we have investigated the prospects of iDM with electromagnetic form factors at the FASER2, MATHUSLA, and SHiP detectors, while we also obtained bounds from previous experiments such as CHARM and NuCal.
We focused on the long-lived regime of the heavier DS species, $\chi_1$, decaying into the lighter DS state and a photon (EDM/MDM portal) or a $e^+ e^-$ pair (AM/CR operator).

In the first case, we extended previous work studying $\chi_1$ decays \cite{Dienes:2023uve} by taking into the account the secondary LLP production occurring at FASER$\nu$2, which will allow FASER2 to cover the $\sim O(1)\,\m$ region of the parameter space.
This sector is particularly significant for MDM iDM, where in a part of this region, $\chi_1$ is a thermal DM candidate.

For the AM/CR operator, we investigated for the first time the main iDM signature in intensity frontier searches, which is the displaced $\chi_1$ decays.
It allows FASER2 to be competitive with LEP bounds, with much greater reach than for the scenario of elastic DS with EM form factors, while SHiP may even cover a part of the allowed parameter space for large values of the mass splitting $\Delta$.

In summary, we have shown that an iDM with electromagnetic form factors leads to greater detection prospects at FASER, MATHUSLA, and SHiP than the elastic coupling case, which was studied previously by DS-electron scattering signatures.
The non-minimal iDM content allows not only for the decays of heavier DS species, but also for upscattering of the lighter one into the heavier one just in front of the decay vessel, which can lead to greater sensitivity than both displaced LLP decays and scattering with electrons.

\acknowledgments

This work was supported by the Institute for Basic Science under the project code, IBS-R018-D1.

\clearpage 

\appendix

\section{Decays of $\chi_1$}
\label{app:decays}

In sections below, we collect the formulas used in our simulations.
In some cases we provided only their leading form.
The full results, including the derivations, can be found in the Mathematica notebook included in \href{https://github.com/krzysztofjjodlowski/Looking_forward_to_photon_coupled_LLPs}{\faGithub}.

Below, we collect the decay widths for two- and three-body decays of $\chi_1$.
For dimension-5 portals, we use results from \cite{Dienes:2023uve}:
\be
  \text{EDM:\quad}   \  
  & \Gamma_{\chi_1 \to \chi_0 \gamma} = \frac{\Delta^3 (\Delta+2)^3 m_{\chi_0}^3}{2 \pi  (\Delta+1)^3 \Lambda_{\mathrm{E}}^2} \simeq \frac{4 \Delta^3 m_{\chi_0}^3}{\pi \Lambda_{\mathrm{E}}^2}, \\
  \text{MDM:\quad}  \  
  & \Gamma_{\chi_1 \to \chi_0 \gamma} = \frac{\Delta^3 (\Delta+2)^3 m_{\chi_0}^3}{2 \pi  (\Delta+1)^3 \Lambda_{\mathrm{M}}^2} \simeq \frac{4 \Delta^3 m_{\chi_0}^3}{\pi \Lambda_{\mathrm{M}}^2}.
  \label{eq:Gamma_EDM_MDM}
\ee

It gives the expression for the photon energy $E_{\chi_0 \gamma}$, which is then used to impose the energy threshold cut.

For dimension-6 portals, such two-body decays are not possible, and the leading decay channel is the three-body decay of $\chi_1$ into $\chi_0$ and a pair of charged leptons.
In the $m_{\chi_0} \gg m_l$ limit, the width of this decay is
\be
  \text{AM:\quad} \     
    & \Gamma_{\chi_1 \to \chi_0 l^+ l^-} \simeq  
        \begin{cases}
          \frac{a_\chi^2 \alpha_{\mathrm{EM}} \Delta^5 m_{\chi_0}^5}{5 \pi ^2}: & \Delta \ll 1, \\
          \frac{a_\chi^2 \alpha_{\mathrm{EM}} m_{\chi_1}^5}{96 \pi ^2}:   & m_{\chi_0}=0,
        \end{cases} \\
  \text{CR:\quad}  \   
    & \Gamma_{\chi_1 \to \chi_0 l^+ l^-} \simeq 
        \begin{cases}
          \frac{b_\chi^2 \alpha_{\mathrm{EM}} \Delta^5 m_{\chi_0}^5}{15 \pi ^2}: & \Delta \ll 1, \\
          \frac{b_\chi^2 \alpha_{\mathrm{EM}} m_{\chi_1}^5}{96 \pi ^2}:   & m_{\chi_0}=0.
        \end{cases}
\label{eq:Gamma_AM_CR}
\ee

The general expression for such decay width, which we use in our simulation, is
\be
  \Gamma_{\chi_1 \to \chi_0 l^+ l^-} =  \int \int \frac{d^2 \Gamma_{\chi_1 \to \chi_0 l^+ l^-}}{d s_{12}d s_{23}} \times \frac{ds_{12} ds_{23}}{\mathrm{Br}(m_{\gamma^*}=\sqrt{s_{23}})},
  \label{eq:Gamma_chi1_tot}
\ee
where $s_{12}=(p_1+p_2)^2$, $s_{23}=(p_2+p_3)^2$, and $p_1$, $p_2$, $p_3$ are the momenta of $\chi_0$, $l^+$, $l^-$, respectively; the first factor is the double differential decay width, and the second one is the branching ratio of a virtual massive photon which accounts for the decays into hadrons taken from the PDG \cite{Workman:2022ynf}.

We also follow PDG kinematics chapter \cite{Workman:2022ynf} for general expression for $\frac{d^2 \Gamma}{d s_{12}d s_{23}}$ and the expressions for the integration limits. 
The amplitude squared averaged over the spins of the leptons reads:
\begin{dmath}[labelprefix={eq:}]
  {|M|_{\text{CR}}^2 = -16 \pi \alpha_{\mathrm{EM}} b_\chi^2} \left(m_{\chi_1}^2 \left(2 m_{\chi_0}^2-2 s_{12}-s_{23}\right) \\
  +2 m_{\chi_1} m_{\chi_0} \left(2 m_l^2+s_{23}\right)-m_{\chi_0}^2 (2 s_{12}+s_{23})+2 \left(m_l^2-s_{12}\right)^2 \\
  +2 s_{12} s_{23}+s_{23}^2\right), \\
  {|M|_{\text{AM}}^2 = -16 \pi \alpha_{\mathrm{EM}}  a_\chi^2} \left(2 \left(-s_{12} \left(m_{\chi_1}^2+m_{\chi_0}^2+2 m_l^2\right) \\
  +\left(m_l^2-m_{\chi_1} m_{\chi_0}\right)^2+s_{12}^2\right)-s_{23} \left((m_{\chi_1}+m_{\chi_0})^2-2 s_{12}\right)+s_{23}^2\right).
\end{dmath}

In the case of three-body, we impose the energy threshold cut in the following way.
For each point in the ($\Lambda$, $m_{\chi_0}$, $\Delta$) parameter space, we calculate the average visible energy of the $e^+ e^-$ pair as a function of $E_{\chi_1}$, 
\begin{dmath}[labelprefix={eq:}]
  \langle E_{e^+ e^-} \rangle (E_{\chi_1}) \equiv \int \int ds_{12} ds_{23} \frac{d^2 \Gamma_{\chi_1 \to \chi_0 e^+ e^-}}{d s_{12}d s_{23}} \times \frac{E_{e^+}+E_{e^-}}{\Gamma_{\chi_1 \to \chi_0 e^+ e^-}},
\end{dmath}
where we use PDG \cite{Workman:2022ynf} for expressions for energies of $e^{\pm}$, $E_{e^\pm} \equiv E_{e^\pm}(E_{\chi_1},s_{12},s_{23})$.
Next, for a given energy threshold, $E_{\mathrm{th}}$, we determine $E_{\chi_1}$ for which $\langle E_{e^+ e^-} \rangle > E_{\mathrm{th}}$. 
Finally, we use this value of $E_{\chi_1}$ as a cut on $\chi_1$.

\section{Pseudoscalar and vector meson decays}
\label{app:prod}

\subsection{Vector meson decays}
It is known that the two-body decays of vector mesons into $\overline{\chi}_0\chi_1$ pair dominate over the three-body pseudoscalar meson decays \cite{Chu:2020ysb,Kling:2022ykt,Dienes:2023uve}.
Below we give the formulas for $V(p_0) \!\to\! \gamma^*(p_1+p_2) \!\to\! \overline{\chi}_0(p_1) + \chi_1(p_2)$, which are mediated by an off-shell photon, 
\begin{widetext}
  \be
    \label{eq:brV}
    \text{EDM:}\quad
      &\frac{{\rm BR}_{V \rightarrow \bar{\chi}_0 \chi_1}}{{\rm BR}_{V \rightarrow e^+ e^-}} \!=\! \frac{ (M^2-(m_{\chi_0}+m_{\chi_1})^2) \left(M^2+2 (m_{\chi_0}-m_{\chi_1})^2\right) \sqrt{\left(-M^2+m_{\chi_0}^2+m_{\chi_1}^2\right)^2-4 m_{\chi_0}^2 m_{\chi_1}^2}}{2 \pi  \alpha \Lambda_{\mathrm{E}}^2 M \sqrt{M^2-4 m_e^2} \left(M^2+2 m_e^2\right)}, \\
    \text{MDM:}\quad
    &\frac{{\rm BR}_{V \rightarrow \bar{\chi}_0 \chi_1}}{{\rm BR}_{V \rightarrow e^+ e^-}} \!=\! \frac{(M^2 - (m_{\chi_0}-m_{\chi_1})^2) \left(M ^2+2 (m_{\chi_0}+m_{\chi_1})^2\right) \sqrt{\left(-M ^2+m_{\chi_0}^2+m_{\chi_1}^2\right)^2-4 m_{\chi_0}^2 m_{\chi_1}^2}}{2\pi\alpha \Lambda_{\mathrm{M}}^2 M  \left(M ^2+2 m_e^2\right) \sqrt{M ^2-4 m_e^2}}, \\
    \text{AM:}\quad
			&\frac{{\rm BR}_{V \rightarrow \bar{\chi}_0 \chi_1}}{{\rm BR}_{V \rightarrow e^+ e^-}} \!=\! \frac{a_\chi^2 M (M^2-(m_{\chi_0}+m_{\chi_1})^2) \left(2 M^2+(m_{\chi_0}-m_{\chi_1})^2\right) \sqrt{\left(-M^2+m_{\chi_0}^2+m_{\chi_1}^2\right)^2-4 m_{\chi_0}^2 m_{\chi_1}^2}}{8 \pi \alpha \sqrt{M^2-4 m_e^2} \left(M^2+2 m_e^2\right)}, \\
    \text{CR:}\quad
			&\frac{{\rm BR}_{V \rightarrow \bar{\chi}_0 \chi_1}}{{\rm BR}_{V \rightarrow e^+ e^-}} \!=\! \frac{b_\chi^2 M (M^2-(m_{\chi_0}-m_{\chi_1})^2) \left(2 M^2+(m_{\chi_0}+m_{\chi_1})^2\right) \sqrt{\left(-M^2+m_{\chi_0}^2+m_{\chi_1}^2\right)^2-4 m_{\chi_0}^2 m_{\chi_1}^2}}{8 \pi \alpha \sqrt{M^2-4 m_e^2} \left(M^2+2 m_e^2\right)},
  \ee
\end{widetext}
where $\rm{BR}_{V \rightarrow e^+ e^-}$ is the branching ratio corresponding to the $V \rightarrow e^+ e^-$ decay, which we took from the PDG \cite{Workman:2022ynf}.

\subsection{Pseudoscalar meson decays}
Below we list the differential branching ratios of the pseudoscalar meson decays into $\gamma$ and $\overline{\chi}_0\chi_1$ pair mediated by an off-shell photon, $P(p_0) \!\to\! \gamma(p_1)+ \gamma^*(p_2+p_3) \!\to\! \gamma(p_1) + \overline{\chi}_0(p_2) + \chi_1(p_3)$,
\begin{widetext}
  \be
    \label{eq:br2dq2dcostheta}
    \text{EDM:}\quad
      \frac{d{\rm BR}_{P \rightarrow \gamma \bar{\chi}_0 \chi_1}}{dq^2 d\cos\theta} = {\rm BR}_{P\rightarrow \gamma \gamma}  &\!\times \!\!  \frac{ \left(q^2-M^2\right)^3 \left(q^2-(m_{\chi_0}+m_{\chi_1})^2\right) \sqrt{\left(m_{\chi_0}^2+m_{\chi_1}^2-q^2\right)^2-4 m_{\chi_0}^2 m_{\chi_1}^2} }{8 \pi ^2 \Lambda_{\mathrm{E}}^2 M^6 q^6} \\
      &\left[-\cos (2 \theta) \left((m_{\chi_0}-m_{\chi_1})^2-q^2\right)-3 (m_{\chi_0}-m_{\chi_1})^2-q^2\right], \\
    \text{MDM:}\quad
    \frac{d{\rm BR}_{P \rightarrow \gamma \bar{\chi}_0 \chi_1}}{dq^2 d\cos\theta} = {\rm BR}_{P\rightarrow \gamma \gamma}  &\!\times \!\! \frac{\left(q^2-M^2\right)^3 \left(q^2-(m_{\chi_0}-m_{\chi_1})^2\right) \sqrt{\left(m_{\chi_0}^2+m_{\chi_1}^2-q^2\right)^2-4 m_{\chi_0}^2 m_{\chi_1}^2} }{8 \pi^2 \Lambda_{\mathrm{M}}^2 M^6 q^6} \\
      &\left[-\cos(2 \theta) \left((m_{\chi_0}+m_{\chi_1})^2-q^2\right)-3 (m_{\chi_0}+m_{\chi_1})^2-q^2 \right], \\
    \text{AM:}\quad
      \frac{d{\rm BR}_{P \rightarrow \gamma \bar{\chi}_0 \chi_1}}{dq^2 d\cos\theta} = {\rm BR}_{P\rightarrow \gamma \gamma}  &\!\times \!\! \frac{a_\chi^2 \left(q^2-M^2\right)^3 \left(q^2-(m_{\chi_0}+m_{\chi_1})^2\right) \sqrt{\left(m_{\chi_0}^2+m_{\chi_1}^2-q^2\right)^2-4 m_{\chi_0}^2 m_{\chi_1}^2} }{32 \pi ^2 M^6 q^4} \\
      & \left[\cos (2 \theta) \left((m_{\chi_0}-m_{\chi_1})^2-q^2\right)-(m_{\chi_0}-m_{\chi_1})^2-3 q^2\right], \\
    \text{CR:}\quad
      \frac{d{\rm BR}_{P \rightarrow \gamma \bar{\chi}_0 \chi_1}}{dq^2 d\cos\theta} = {\rm BR}_{P\rightarrow \gamma \gamma}  &\!\times \!\! \frac{b_\chi^2 \left(q^2-M^2\right)^3 \left(q^2 - (m_{\chi_0}-m_{\chi_1})^2\right) \sqrt{\left(m_{\chi_0}^2+m_{\chi_1}^2-q^2\right)^2-4 m_{\chi_0}^2 m_{\chi_1}^2}}{32 \pi ^2 M^6 q^4}\\
      & \left[\cos (2 \theta) \left((m_{\chi_0}+m_{\chi_1})^2-q^2\right)-(m_{\chi_0}+m_{\chi_1})^2-3 q^2\right],
  \ee
\end{widetext}
This form of the differential branching ratio allows for straightforward MC simulation of the pseudoscalar meson decays and is the form used in \texttt{FORESEE}.

\section{Cross sections for electron or Primakoff upscattering}
\label{app:e_scat}

Below, we give the leading form of the expressions for the upscattering process, $\chi_0 T \to \chi_1 T$, where the target is either an electron or a nucleus, $T=\{e^-, N\}$.

\subsection{Electron upscattering}

\begin{dmath}[labelprefix={eq:}]
  \text{EDM:\quad}   
    {\frac{d \sigma_{\chi_0 e^- \to \chi_1 e^-}}{d E_R} \simeq \frac{4 \alpha_{\mathrm{EM}}}{\Lambda_{\mathrm{E}}^2 E_{\chi_0}^2} \times} \left(- \frac{m_{\chi_0}^2}{2 m_e} +  \frac{E_{\chi_0}^2}{E_R} -\Delta \frac{E_{\chi_0} m_{\chi_0}^2}{E_R m_e} \right),\\
  \text{MDM:\quad}  
    {\frac{d \sigma_{\chi_0 e^- \to \chi_1 e^-}}{d E_R} \simeq \frac{4 \alpha_{\mathrm{EM}}}{\Lambda_{\mathrm{M}}^2 E_{\chi_0}^2} \times} \left( \frac{m_{\chi_0}^2}{2 m_e} +  \frac{E_{\chi_0}^2}{E_R} -\Delta \frac{E_{\chi_0} m_{\chi_0}^2}{E_R m_e} \right), \\
  \text{AM:\quad}
  {\frac{d \sigma_{\chi_0 e^- \to \chi_1 e^-}}{d E_R} \simeq \frac{\alpha_{\mathrm{EM}} a_\chi^2}{E_{\chi_0}^2} \times} \left[ 2 m_e  E_{\chi_0}^2 + E_R  (m_{\chi_0}^2 - 2 E_{\chi_0} m_e )  - 2 \Delta E_{\chi_0} m_{\chi_0}^2 \right], \\
  \text{CR:\quad}
      {\frac{d \sigma_{\chi_0 e^- \to \chi_1 e^-}}{d E_R} \simeq \frac{\alpha_{\mathrm{EM}} b_\chi^2}{E_{\chi_0}^2} \times} \left[ 2 m_e  E_{\chi_0}^2 - E_R  (m_{\chi_0}^2 + 2 E_{\chi_0} m_e )  - 2 \Delta E_{\chi_0} m_{\chi_0}^2 \right],
  \label{eq:Prim_e_dsigmadER}
\end{dmath}
where $E_R=E_{e^-}-m_{e^-}$ is the recoil energy of the electron.
We have checked that in the $\Delta \to 0$ limit we reproduce the elastic scattering results of \cite{Chu:2018qrm}.

The above expressions can be actually integrated analytically which results in
\be
  \text{EDM:\quad}   \  
    & \sigma_{\chi_0 e^- \to \chi_1 e^-} \simeq \frac{4 \alpha_{\mathrm{EM}}}{\Lambda_{\mathrm{E}}^2} \log\left(\frac{E_R^{\mathrm{max}}}{E_R^{\mathrm{min}}}\right), \\
  \text{MDM:\quad}  \
  & \sigma_{\chi_0 e^- \to \chi_1 e^-} \simeq \frac{4 \alpha_{\mathrm{EM}}}{\Lambda_{\mathrm{M}}^2} \log\left(\frac{E_R^{\mathrm{max}}}{E_R^{\mathrm{min}}}\right), \\
  \text{AM:\quad} \    
    & \sigma_{\chi_0 e^- \to \chi_1 e^-} \simeq 2 a_\chi^2 \alpha_{\mathrm{EM}} m_e (E_R^{\mathrm{max}}-E_R^{\mathrm{min}}), \\
  \text{CR:\quad}  \   
    & \sigma_{\chi_0 e^- \to \chi_1 e^-} \simeq 2 b_\chi^2 \alpha_{\mathrm{EM}} m_e (E_R^{\mathrm{max}}-E_R^{\mathrm{min}}).
\label{eq:Prim_e_integr}
\ee

\section{Thermally averaged annihilation cross sections}
\label{app:relic}

Below, we give all the the thermally averaged $2 \to 2$ cross sections entering into \cref{eq:boltzmann_three}.
These are needed to determine the relic density of $\chi_0$, as described in \cref{sec:relic_density}.
We note that in $\Delta \to 0$ limit we recover results of \cite{Chu:2018qrm}, when it is relevant.

Co-annihilation into charged leptons is given by
\begin{align}
  \langle \sigma v\rangle_{\chi_0\chi_1 \to l^+ l^-} =& \sqrt{(m_{\chi_0} + m_{\chi_1})^2-4 m_l^2} \nonumber\\
   & \times \frac{\left((m_{\chi_0} + m_{\chi_1})^2+2 m_l^2\right)}{m_{\chi_0}+m_{\chi_1}} \times C,
   \label{eq:sigmav_ll}
\end{align}
where the coefficient $C$ is
\be
\label{eq:sigmav_ll_C}
  \text{EDM:\quad}  \
   &\ \  C = \frac{\alpha_{\mathrm{EM}} \left(3 m_{\chi_0}^2-2 m_{\chi_0} m_{\chi_1}+3 m_{\chi_1}^2\right)}{3 \Lambda_{\mathrm{E}}^2 (m_{\chi_0}+m_{\chi_1})^4} \langle v_{\rm rel}^2\rangle
   \,, \\ 
  \text{MDM:\quad}  \
  &\ \  C = \frac{4 \alpha_{\mathrm{EM}}}{\Lambda_{\mathrm{M}}^2 (m_{\chi_0}+m_{\chi_1})^2}
  \,, \\
  \text{AM:\quad}  \
   &\ \  C = \frac{a_\chi^2 \alpha_{\mathrm{EM}} \left(3 m_{\chi_0}^2+2 m_{\chi_0} m_{\chi_1}+3 m_{\chi_1}^2\right)}{12 (m_{\chi_0}+m_{\chi_1})^2} \langle v_{\rm rel}^2\rangle
   \,, \\
   \text{CR:\quad}  \
   &\ \  C = \alpha_{\mathrm{EM}} b_\chi^2
\,,
\ee
and where
\be
  \langle v_{\rm rel}^2\rangle =3T\left(\frac{1}{m_{\chi_0}}+\frac{1}{m_{\chi_1}}\right)
\,.
\ee

Following \cite{Chu:2018qrm}, we take into account co-annihilations into hadrons according to the following formula:
\be
  \langle \sigma v\rangle_{\chi_0\chi_1 \to \mathrm{hadrons}}(s) = \langle \sigma v\rangle_{\chi_0\chi_1 \to \mu^+ \mu^-}(s) \times R(\sqrt{s}),
  \label{eq:sigmav_hadrons}
\ee
where we use the data from PDG \cite{Workman:2022ynf} for the experimentally measured $R$-ratio.

The annihilations into a photon pair is given by
\be
  \langle \sigma v\rangle_{\chi_0\chi_0 \to \gamma \gamma} = \frac{16 m_{\chi_0}^4 m_{\chi_1}^2}{\pi  \left(m_{\chi_0}^2+m_{\chi_1}^2\right)^2} \times C,
  \label{eq:sigmav_gammagamma}
\ee
with
\be
  \text{EDM:\quad}  \
  &\ \  C = \frac{1}{\Lambda_{\mathrm{E}}^4}, \\
  \text{MDM:\quad}  \
  &\ \  C = \frac{1}{\Lambda_{\mathrm{M}}^4}, \\
  \text{AM:\quad}  \
   &\ \  C = 0, \\
   \text{CR:\quad}  \
   &\ \  C = 0,
   \label{eq:sigmav_gammagamma_C}
\ee
and the formula for $\langle \sigma v\rangle_{\chi_1\chi_1 \to \gamma \gamma}$ is obtained from the above by replacing each $m_{\chi_0}$ with $m_{\chi_1}$ and vice versa.

The $\chi_1{\text -}\chi_0$ conversion processes are determined by pair-annihilation
\be
  \langle \sigma v\rangle_{\chi_1\chi_1 \to \chi_0\chi_0} = \frac{4 \sqrt{2} \sqrt{\Delta} m_{\chi_0}^2}{\pi } \times C,
  \label{eq:sigmav_chi1chi1_chi0chi0}
  \ee
  where
  \be
  \text{EDM/MDM:\quad}  \
  &\ \  C = \frac{2}{\Lambda_{\mathrm{E/M}}^4}, \\
  \text{AM:\quad}  \
  &\ \  C = a_\chi^4 \Delta^2 m_{\chi_0}^4, \\
  \text{CR:\quad}  \
  &\ \  C = b_\chi^4 \Delta^2 m_{\chi_0}^4,
  \label{eq:sigmav_chi1chi1_chi0chi0_C}
\ee
and by upscattering
\be
  \langle \sigma v\rangle_{\chi_1 e^- \to \chi_0 e^-} = C,
  \label{eq:sigmav_chi1e_chi0e}
\ee
with
\be
  \text{EDM:\quad}  \
    &\ \  C = \frac{8\sqrt{2} \alpha_{\mathrm{EM}} \sqrt{\frac{m_e}{\Delta m_{\chi_0}}}}{\Lambda_{\mathrm{E}}^2}, \\
  \text{MDM:\quad}  \
  &\ \  C = \frac{8 \sqrt{2} \alpha_{\mathrm{EM}} \sqrt{\frac{\Delta m_{\chi_0}}{m_e}}}{\Lambda_{\mathrm{M}}^2}, \\
  \text{AM:\quad}  \
    &\ \  C = 6\sqrt{2} a_\chi^2 \alpha_{\mathrm{EM}} \Delta m_{\chi_0} \sqrt{\Delta m_{\chi_0} m_e}, \\
  \text{CR:\quad}  \
    &\ \  C = 4 \sqrt{2} \alpha_{\mathrm{EM}} b_\chi^2 m_e \sqrt{\Delta m_{\chi_0} m_e}.
    \label{eq:sigmav_chi1e_chi0e_C}
\ee

\bibliography{main}

\end{document}